\documentclass[
aps,
singlecolumn,
superscriptaddress,
longbibliography]{revtex4-1}  

\usepackage{amssymb}
\usepackage[usenames,dvipsnames]{xcolor}
\usepackage{tikz}
\usetikzlibrary{arrows.meta}
\tikzset{every picture/.style={line width=1pt}}
\usepackage{amsmath}
\usepackage{amssymb}
\usepackage{graphicx}
\usepackage{pgfplots}
\pgfplotsset{compat = 1.3}
\usepackage{soul}

\newcommand{\rhol}{\rho}
\newcommand{\rhosheet}{\rho_e}
\newcommand{\rhosphere}{\rho_s}
\newcommand{\rs}{R_s}
\newcommand{\rf}{R_f}
\newcommand{\Rf}{\rf}
\newcommand{\rc}{r_c}
\newcommand{\lstar}{\ell_*}

\newcommand{\h}{h}
\newcommand{\V}{V}
\newcommand{\epsrr}{\epsilon_{rr}}
\newcommand{\epstt}{\epsilon_{\theta \theta}}
\newcommand{\epsrt}{\epsilon_{r \theta}}
\newcommand{\srrf}{\sigma_{rr}}
\newcommand{\sttf}{\sigma_{\theta \theta}}
\newcommand{\srtf}{\sigma_{r \theta}}
\newcommand{\spar}{\delta}

\newcommand{\as}[2]{{ \ensuremath{ {#1}^{({#2})} } }}

\newcommand{\R}[1]{\ensuremath{ \hat{#1}  } }
  
\newcommand{\w}[1]{ \ensuremath{ \as{w}{#1} } }
\newcommand{\bw}[1]{ \ensuremath{ \as{\bar{w}}{#1} }  }
\newcommand{\p}[1]{ \as{\varphi}{#1} }
\newcommand{\bp}[1]{ \as{\bar{\varphi}}{#1} }
\newcommand{\ur}[1]{ \as{u_r}{#1} }

\newcommand{\bur}[1]{ \as{\bar{u}_r}{#1} }
\newcommand{\but}[1]{ \as{\bar{u}_{\theta} }{#1} }
\newcommand{\stt}[1]{\as{\sigma_{\theta \theta}}{#1} }

\newcommand{\sttb}[1]{\as{\bar{\sigma}_{\theta \theta}}{#1} }

\newcommand{\btheta}{\tilde{\theta}}
\newcommand{\bz}{\tilde{z}}
\newcommand{\bt}{\tilde{t}}

\newcommand{\beq}{\begin{equation}}
\newcommand{\eeq}{\end{equation}}

\newcommand{\rca}{\rc^{(0)}}

\newcommand{\lcurv}{\ell_{\mathrm{curv}}}
\newcommand{\lc}{\ell_c}

\newcommand{\wno}{k}
\newcommand{\wnostar}{\wno_*}
\newcommand{\wnob}{\kappa}
\newcommand{\coeff}{a}
\newcommand{\simpar}{\alpha}

\newcommand{\Ai}{\mbox{Ai}}

\newcommand{\con}{\xi^*}

\newcommand{\f}{f_0}
\newcommand{\Hsim}{H_0}
\newcommand{\fcor}{f_1}
\newcommand{\Hcor}{H_1}
\newcommand{\ff}{f}
\newcommand{\Hsimf}{H}

\newcommand{\unitk}{\hat{\boldsymbol{k}}}

\newcommand{\W}{W}
\newcommand{\Pw}{\Phi}
\newcommand{\C}{P}

\newcommand{\depth}{\Delta}

\newcommand{\lin}{L_{I}}
\newcommand{\lout}{L_{O}}

\newcommand{\sigmastar}{\sigma_{*}^{rr}}

\newcommand{\weber}{\operatorname{We}}
\newcommand{\lweber}{L}

\newcommand{\x}{x}

\definecolor{col1}{rgb}{0.2081, 0.1663, 0.5292}
\definecolor{col2}{rgb}{0.2124, 0.2116, 0.6224}
\definecolor{col3}{rgb}{0.1989, 0.2597, 0.7186}
\definecolor{col4}{rgb}{0.1403, 0.3147, 0.8168}
\definecolor{col5}{rgb}{0.0224, 0.3786, 0.8793}
\definecolor{col6}{rgb}{0.0106, 0.4187, 0.8809}
\definecolor{col7}{rgb}{0.0410, 0.4502, 0.8685}
\definecolor{col8}{rgb}{0.0668, 0.4791, 0.8523}
\definecolor{col9}{rgb}{0.0788, 0.5082, 0.8362}
\definecolor{col10}{rgb}{0.0734, 0.5410, 0.8257}
\definecolor{col11}{rgb}{0.0473, 0.5790, 0.8227}
\definecolor{col12}{rgb}{0.0265, 0.6137, 0.8135}
\definecolor{col13}{rgb}{0.0232, 0.6407, 0.7925}
\definecolor{col14}{rgb}{0.0254, 0.6623, 0.7637}
\definecolor{col15}{rgb}{0.0527, 0.6812, 0.7305}
\definecolor{col16}{rgb}{0.1024, 0.6984, 0.6934}
\definecolor{col17}{rgb}{0.1640, 0.7141, 0.6527}
\definecolor{col18}{rgb}{0.2360, 0.7281, 0.6086}
\definecolor{col19}{rgb}{0.3187, 0.7395, 0.5625}
\definecolor{col20}{rgb}{0.4082, 0.7466, 0.5185}
\definecolor{col21}{rgb}{0.4951, 0.7491, 0.4806}
\definecolor{col22}{rgb}{0.5745, 0.7484, 0.4479}
\definecolor{col23}{rgb}{0.6473, 0.7456, 0.4188}
\definecolor{col24}{rgb}{0.7153, 0.7414, 0.3917}
\definecolor{col25}{rgb}{0.7798, 0.7361, 0.3658}
\definecolor{col26}{rgb}{0.8419, 0.7307, 0.3398}
\definecolor{col27}{rgb}{0.9025, 0.7262, 0.31215}
\definecolor{col28}{rgb}{0.9613, 0.7281, 0.2774}
\definecolor{col29}{rgb}{0.9978, 0.7542, 0.2289}
\definecolor{col30}{rgb}{0.9931, 0.7936, 0.1901}
\definecolor{col31}{rgb}{0.9763, 0.8328, 0.1590}
\definecolor{col32}{rgb}{0.9616, 0.8748, 0.1278}
\definecolor{col33}{rgb}{0.9602, 0.9244, 0.0931}
\definecolor{col33}{rgb}{0.9763, 0.9831, 0.0538}    

\definecolor{val1}{rgb}{0.2422 0.1504 0.6603}
\definecolor{val2}{rgb}{0.2702 0.2135 0.8340}
\definecolor{val3}{rgb}{0.2813 0.2978 0.9391}
\definecolor{val4}{rgb}{0.2713 0.3850 0.9895}
\definecolor{val5}{rgb}{0.2021 0.4788 0.9911}
\definecolor{val6}{rgb}{0.1724 0.5639 0.9405}
\definecolor{val7}{rgb}{0.1300 0.6391 0.8918}
\definecolor{val8}{rgb}{0.0478 0.7025 0.8267}
\definecolor{val9}{rgb}{0.0704 0.7457 0.7258}
\definecolor{val10}{rgb}{0.2028 0.7796 0.6111}
\definecolor{val11}{rgb}{0.3594 0.8018 0.4629}
\definecolor{val12}{rgb}{0.5791 0.7924 0.2907}
\definecolor{val13}{rgb}{0.7858 0.7573 0.1598}
\definecolor{val14}{rgb}{0.9456 0.7285 0.2167}
\definecolor{val15}{rgb}{0.9948 0.7922 0.2007}
\definecolor{val16}{rgb}{0.9608 0.8906 0.1530}
\definecolor{val17}{rgb}{0.9769 0.9839 0.0805}

\definecolor{c1}{rgb}{0.2422    0.1504    0.6603}
\definecolor{c2}{rgb}{0.2760    0.2395    0.8785}
\definecolor{c3}{rgb}{0.2780    0.3556    0.9777}
\definecolor{c4}{rgb}{0.2021    0.4788    0.9911}
\definecolor{c5}{rgb}{0.1540    0.5902    0.9218}
\definecolor{c6}{rgb}{0.0899    0.6837    0.8550}
\definecolor{c7}{rgb}{0.0704    0.7457    0.7258}
\definecolor{c8}{rgb}{0.2380    0.7901    0.5658}
\definecolor{c9}{rgb}{0.5044    0.7993    0.3480}
\definecolor{c10}{rgb}{0.7858    0.7573    0.1598}
\definecolor{c11}{rgb}{0.9871    0.7348    0.2438}
\definecolor{c12}{rgb}{0.9700    0.8576    0.1673}
\definecolor{c13}{rgb}{0.9769    0.9839    0.0805}

\begin{document}

\title{ Impact on floating thin elastic sheets: A mathematical model}

\author{Doireann O'Kiely}%
\email{okiely@maths.ox.ac.uk}
\affiliation{Mathematical Institute, University of Oxford, Oxford OX2 6GG, United Kingdom}
\author{Finn Box}
\affiliation{Mathematical Institute, University of Oxford, Oxford OX2 6GG, United Kingdom}
 \author{Ousmane Kodio}
 \affiliation{Mathematical Institute, University of Oxford, Oxford OX2 6GG, United Kingdom}
  \author{Jonathan Whiteley}
 \affiliation{Department of Computer Science, University of Oxford, Oxford OX1 3QD, United Kingdom}
  \author{Dominic Vella}
\email{dominic.vella@maths.ox.ac.uk}
  \affiliation{Mathematical Institute, University of Oxford, Oxford OX2 6GG, United Kingdom}

\begin{abstract}
We investigate impact of a sphere onto a floating elastic sheet and the resulting formation and evolution of wrinkles in the sheet.
Following impact, we observe a radially propagating wave,  beyond which the sheet remains approximately planar but is decorated by a series of radial wrinkles whose wavelength grows in time.  We develop a mathematical model to describe these phenomena by exploiting the asymptotic limit in which the bending stiffness is small compared to stresses in the sheet.  The results of this analysis show that, at a time $t$ after impact, the transverse wave is located at a radial distance $r\sim t^{1/2}$ from the impactor, in contrast to the classic $r\sim t^{2/3}$ scaling observed for  capillary--inertia ripples produced by dropping a stone into a pond.  We describe the shape of this wave, starting from the simplest case of a point impactor, but subsequently addressing a finite-radius spherical impactor, contrasting this case with the classic Wagner theory of impact.  We show also that the coarsening of wrinkles in the flat portion of the sheet is controlled by the inertia of the underlying liquid: short-wavelength, small-amplitude wrinkles form at early times since they accommodate the geometrically-imposed compression without significantly displacing the underlying liquid. As time progresses, the liquid accelerates and the wrinkles grow larger and coarsen.  We explain this  coarsening quantitatively using numerical simulations and scaling arguments, and we compare our predictions with experimental data.
\end{abstract}

\makeatother
\maketitle

\section{Introduction}
When a solid impacts  a liquid--air interface, the response varies from dramatic deceleration of the solid by hydrodynamic forces (in the case of ship slamming~\cite{Howison1991, Philippi2016}), to the more peaceful `plop' heard when a pebble is dropped into a pond and the resulting cavity  collapses \cite{Worthington1900,Birkhoff1957}. 
The large hydrodynamic forces that are generated during such impacts allow animals to run on the surface of water \cite{BushHu2006}, provided that they can move their feet out of the cavity before it collapses  \cite{Glasheen1996a,Glasheen1996b,Gaudet1998}. While  the splashes induced by such impacts has proved a fertile area for research, various strategies have been proposed for avoiding splash, including modifying the coating of the impactor \cite{Duez2007}. As an alternative to modifying the impactor, the addition of a thin elastic sheet at the liquid--air interface also changes its behaviour and has been proposed as a way of modifying (or reducing) splash \cite{poopsplash,Pepper2008,Gielen2017}. 

The mechanics of very thin elastic sheets, and their interaction with liquid--air interfaces, is a subject of considerable interest in its own right. Such sheets exhibit deviations from the classic Young--Dupr\'{e} equation for the contact angle of a drop  \cite{Schulman2015,Nadermann2013} and may wrinkle in response to capillary forces  \cite{Huang2007,Schroll2013}, or localized indentation \cite{Vella2015PRL,Paulsen2016pnas,Box2017}. Previous studies have largely focused on static scenarios in which the state of the system, including the number of wrinkles, may be determined by equilibrium arguments such as energy minimization. Some dynamic experiments \cite{BugraToga2013} have attempted to resolve the propagation of wrinkles as the equilibrium state is approached, and suggest that the number of wrinkles reaches its equilibrium value before the wrinkles have grown to their equilibrium length. This  appears to conflict with the standard equilibrium picture in which the wrinkle length is determined by the minimization of a dominant energy while the number of wrinkles is determined by the minimization of a sub-dominant energy \cite{Davidovitch2011}.

Previous work on dynamic impact-induced wrinkling has focused on situations in which the sheets are relatively extensible: Vermorel \emph{et al.}~\cite{VVV2009} studied a sheet with zero applied background stress impacted by a sphere, which shows the propagation of a longitudinal sound wave in the sheet followed by wrinkling. More recently, impact onto a floating sheet has been considered in two-dimensional~\cite{Duchemin2014} and axisymmetric \cite{Vandenberghe2016} configurations. Again, these experiments demonstrated that a tensile wave emanates from the point of impact and  travels through the sheet at the speed of sound. This is followed by a transverse wave and, in axisymmetric sheets, the formation of wrinkles as the sheet is compressed in the azimuthal direction. However, in these scenarios the stretching of the elastic sheet was significant and the in-plane stresses were not in quasistatic equilibrium.

In this paper, we study the dynamic response  to impact of a very thin elastic sheet floating at a liquid--air interface, following experiments presented elsewhere \cite{pnas}. These sheets have very small resistance to bending (in a way that we shall make precise in due course), which leads to significant modifications of previous work in  the fluid dynamics of impact, while the dynamics of impact change the wrinkle phenomenology compared to previous static work.  We begin in \S\ref{sec:background} by describing the important related research areas: we summarize results from previous studies on static wrinkling as well as impact at a liquid--air interface (which will both serve as comparisons with the dynamic scenario studied here) and then give an overview of experimental observations of impact on a floating elastic sheet.
In \S\ref{sec:intro}, we develop a dimensionless model for the response of a thin elastic sheet floating on an inviscid fluid to a point impactor moving at constant velocity.   We consider the asymptotic limit in which the sheet provides minimal resistance to bending, and the wrinkle wavelength is very small.  In \S\ref{sec:elastocapillarywave}, we derive a simplified model for the transverse wave that propagates radially in the sheet.  We determine the self-similar shape adopted by the floating sheet at early times, and estimate the radial retraction of the sheet that this deformation causes.   In \S\ref{sec:finitealpha} we extend our model to account for finite impactor size, spanning between the point and large impactor limits.  This allows us to compare the predictions of our model quantiatively with experimental observations.
In \S\ref{sec:wrinkles}, we focus on wrinkle evolution: we find that the inertia of the underlying fluid dictates that the wrinkles are initially very fine, but then gradually coarsen, while a geometric constraint causes deviation from the power-law coarsening one might naively expect.  We investigate this through the numerical solution of a simple model governing the amplitude of different wrinkle modes that results from azimuthal compression.   In \S\ref{sec:comparison} we discuss the dynamic behaviour observed here in the context of previous studies on static wrinkling patterns as well as viscosity-dominated wrinkle coarsening. Finally, in \S\ref{sec:discussion}, we summarize our findings and discuss areas that require further investigation.  

\section{Relevant physical phenomena} \label{sec:background}
\subsection{Static wrinkling: Summary}\label{sec:static}
Numerous studies have examined static wrinkles in  floating elastic sheets that are indented so that their centre lies a depth $\depth$ below the free surface \cite{Holmes2010,Vella2015PRL,Paulsen2016pnas,Box2017,Vella2018,Ripp2018_arxiv}.  
As the indentation depth $\depth$ is increased quasistatically, the sheet transitions from unwrinkled to wrinkled, and the wrinkles then evolve as $\depth$ is increased further --- here we give a brief overview of the relevant results for ease of later comparison; the interested reader is referred to ref.~\cite{Vella2018} for a more detailed discussion.

For very small indentation depths, the sheet experiences approximately uniform, isotropic tension and vertical deflections are limited to a horizontal length comparable to the capillary length $\lc=\sqrt{\gamma/\rho g}$, where $\gamma$ is the surface tension of the liquid--air interface, $\rho$ is the density of the liquid and $g$ is the acceleration due to gravity.  The vertical deflection of the sheet pulls material circles radially inwards, which in turn leads to a relative compression in the hoop (azimuthal) direction: the perimeter of a material circle is too large for its new position, and it must therefore compress slightly. This relative compression increases as the indentation depth increases. Beyond a critical indentation depth, the radial retraction of the sheet is large enough that the sheet experiences absolute compression in the azimuthal direction.  If the floating sheet is very thin (or highly `bendable'), then it is unable to sustain a compressive stress and instead forms radial wrinkles, which in turn alter the stress profile in the sheet. In particular, where the sheet is wrinkled, the azimuthal stress $\sigma_{\theta\theta} \approx 0$ \cite{Davidovitch2011}, which combined with in-plane equilibrium ($\nabla\cdot\boldsymbol{\sigma}=0$) gives a radial tension $\sigma_{rr} \approx C/r$ for some constant $C$ (which in general depends on $\depth$). Wrinkles initially form within an annulus away from both the indenter and the edge of the sheet $r = \rf$, but as the indentation depth increases they rapidly reach the edge of the sheet \cite{Vella2018}.  This occurs at an indentation depth $\depth \sim (\gamma/E\h)^{1/2} \ell_c^{1/3} R_f^{2/3}$, and beyond this, the constant $C=\gamma\rf$ for the remainder of the experiment.  The equilibrium vertical deflection  of the sheet, $w(r)$, is then~\cite{Vella2015PRL, Paulsen2016pnas,Vella2018}
\begin{equation}
w(r) \approx \depth \frac{ \Ai \left( r/\lcurv\right)}{\Ai(0)},
\end{equation}
where $\Ai(x)$ is the  Airy function that decays in the far field and
\begin{equation}
\lcurv = \rf^{1/3}\lc^{2/3}
\label{eqn:lcurvStatic}
\end{equation} 
is the  horizontal lengthscale over which the sheet is curved; for $r\gtrsim\lcurv$ the sheet is flat. The new lengthscale $\lcurv$ depends on the sheet radius as well as the classical capillary length $\ell_c = (\gamma/\rho g)^{1/2}$.  Like the classical capillary length, $\lcurv$ corresponds to a balance between the hydrostatic pressure in the fluid and the pressure jump across the interface, but is modified because the interfacial tension now varies radially as a result of wrinkling.  We note that, once wrinkles have reached the edge of the sheet, this horizontal lengthscale of the vertical deflection is constant and is not affected by further increases in indentation depth.  

The microscopic structure of the wrinkle pattern is governed by a balance between the bending stiffness of the elastic sheet (which prefers larger wavelengths) and a resisting stiffness (which prefers smaller wavelengths)~\cite{CerdaMahadevan2003}.  For a sheet of bending stiffness $B = E \h^3/[12(1-\nu^2)]$ (with $E$ its Young's modulus, $\h$ its thickness and $\nu$ its Poisson ratio), there is a locally energetically favourable wavelength $\lambda$ (in the absence of tension) given by~\cite{Paulsen2016pnas}
\begin{equation}
\lambda = 2\pi \left( \frac{B}{K} \right)^{1/4},
\label{eqn:LambdaStatic}
\end{equation}
where $K$ denotes an effective substrate stiffness which may be simply the true substrate stiffness or may be augmented by curvature along the wrinkles~\cite{CerdaMahadevan2003, Paulsen2016pnas}.  The situation becomes more complicated when the geometry of the system necessitates the nucleation of new wrinkles to attain the locally favourable $\lambda$, or when tension along the wrinkles introduces a new energetic cost, but we will not address these complications here.  For quasistatic indentation of a floating sheet, a curvature-induced stiffness dominates the curved part of the sheet so that $K\sim Eh(w'')^2$~\cite{Paulsen2016pnas} and the number of wrinkles $m \sim \sqrt{\Delta/h}$ increases as $\Delta$ increases.  By contrast, in the flat portion of the sheet, hydrostatic pressure within the liquid dominates the substrate's response so that $K = \rho g$ and the number of wrinkles (and hence the wrinkle wavelength) does not change with increasing confinement.

We shall see that the two key features of static wrinkling mentioned in this brief review (namely a constant horizontal lengthscale $\lcurv=\rf^{1/3}\lc^{2/3}$ and fixed wrinkle wavelength in the flat portion of the sheet) are qualitatively different when wrinkling is induced by impact.

\subsection{Hydrodynamic impact: summary} \label{sec:bareimpact}
In this section we summarize the main results for impact at a bare liquid--air interface.  We focus on the case where gravity is unimportant, since that is the limit we will consider below for impact on a floating elastic sheet.  When an object impacts at a liquid--air interface, it imparts momentum to the fluid, inducing a flow.  For an inviscid irrotational fluid, this flow can be described in terms of a velocity potential $\phi$, with the fluid velocity $\boldsymbol{v} = \nabla \phi$; if the fluid is further assumed to be incompressible then the velocity potential is governed by $\nabla^2 \phi = 0$ in the bulk.  The pressure $p$ associated with this flow must satisfy an appropriate condition at the liquid interface, yielding a Bernoulli equation of the form
\begin{equation} \label{eq:sumbernoulli}
p = -\rho \left( \frac{\partial \phi}{\partial t} + \frac{1}{2} | \nabla \phi |^2 \right)
\end{equation}
at the interface.  Surface tension at the liquid--air interface induces a pressure jump across that interface, with
\begin{equation}
[p]^+_- = \gamma \kappa,
\end{equation}
where $\gamma$ is the surface tension coefficient and $\kappa$ the curvature of the surface.  We can now consider two limiting cases, corresponding to small or large values of the Weber number $\weber = \rho V^2 \lweber/\gamma$, where $V$ is the impact speed and $\lweber$ is the size of the impactor. When $\weber \ll 1$,  the velocity and pressure induced by the impact are relatively small and surface tension makes a significant contribution to the pressure, while for $\weber \gg 1$ the velocity and pressure associated with impact are relatively large, and the effect of surface tension is negligible.

In the limit $\weber \ll 1$, the disturbance in the fluid is moderated by surface tension at the interface, leading to so-called capillary waves emanating from the impact point.  At early times when fluid velocities are small, the nonlinear term in~\eqref{eq:sumbernoulli} may be neglected, and the hydrodynamic pressure $p \approx - \rho \partial \phi/\partial t$ balances the pressure jump due to surface tension.  To determine the curvature of the interface and close the problem, we impose a kinematic boundary condition $\partial w/\partial t = \partial \phi/\partial z$ relating the motion of the interface (which is at height $w$) to the fluid velocity.  A straightforward scaling argument then yields the result that the wavefront $r_m$ will propagate according to $r_m \sim \left(\gamma t^2/\rho\right)^{1/3}$.  
This scaling was proposed by Keller and Miksis~\cite{KellerMiksis1983} and appears in many surface tension driven flows (see e.g.~\cite{KellerMiksis1983, BillinghamKing1995, VellaMetcalfe2007}).

By contrast, if $\weber \gg 1$ then surface tension is negligible, but the fluid at the interface must move to conform to the shape of the impactor.  For small deadrise angles, the leading-order hydrodynamic pressure is again $p \approx - \rho \partial \phi/\partial t$~\cite{JimThesis}.  Provided the impactor is smooth, the shape of the contact region can be approximated by a surface height $w = -t + r^2/2R_s$, where $R_s$ is a constant dependent on the curvature of the impactor.  This yields a different self-similar response with radial extent $r \sim t^{1/2}$ introduced by Philippi \emph{et al.}~\cite{Philippi2016}.  This model is appropriate for systems such as ship slamming where the impactor size and velocity are very large, and dates back to the work of Wagner~\cite{Wagner1932}.

\subsection{Dynamic wrinkling: phenomenology}
\begin{figure}[ht]
\centering
\begin{tikzpicture}
\node at (0,0) {  \includegraphics[width=16cm]{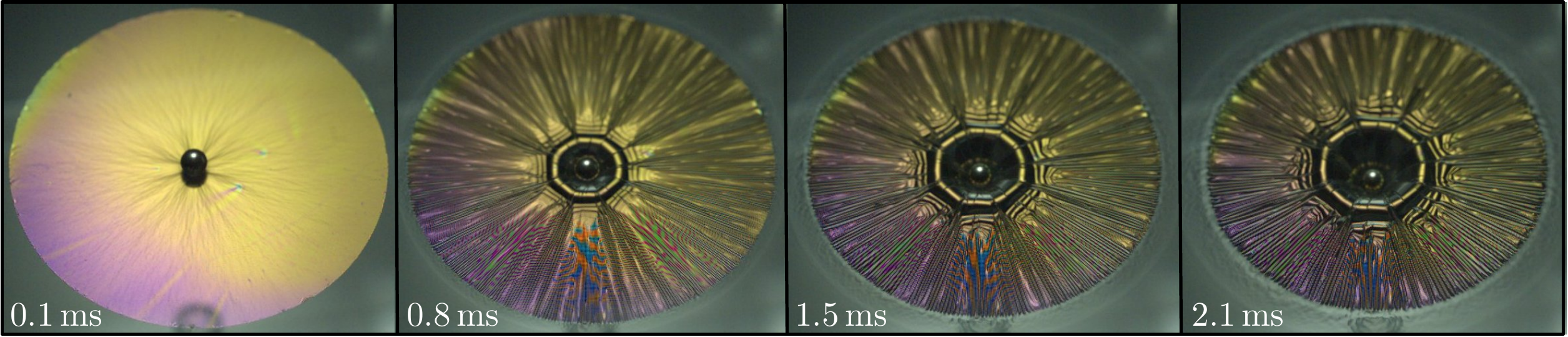} 
};
\node at (-7.6,2) {(a)};
\node at (-3.6,2) {(b)};
\node at (+0.4,2) {(c)};
\node at (+4.4,2) {(d)};
\end{tikzpicture}
\caption{Snapshots of a steel sphere, of radius $R_s = 1.25$\,mm and mass $0.063$\,g, impacting a floating polystyrene sheet (of radius $\Rf= 17.46\mathrm{~mm}$ and thickness $\h = 350\mathrm{~nm}$) at a vertical speed $V=0.69\mathrm{~m/s}$. Images are shown at times $t=0.1$, 0.8, 1.5 and 2.1\,ms after impact.}
\label{fig:snapshots}
\end{figure}
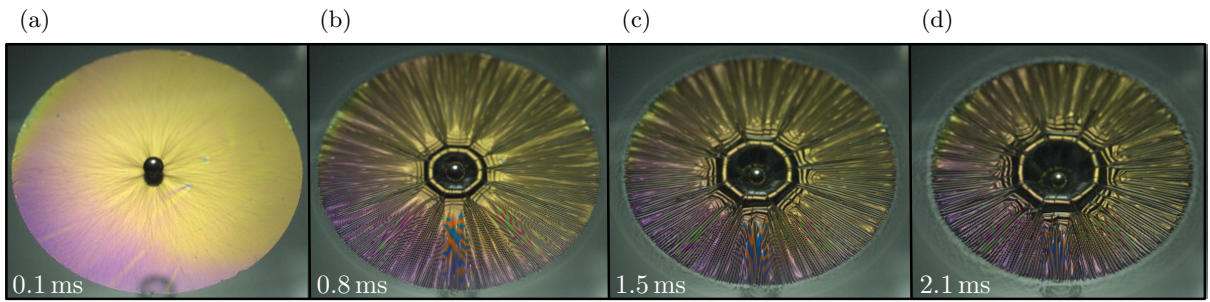

A related study~\cite{pnas} presented  experiments in which thin polystyrene sheets (thickness in the range $150\mathrm{~nm}\leq \h\leq530\mathrm{~nm}$, radius in the range $5.7\mathrm{~mm}\leq\Rf\leq17.7\mathrm{~mm}$) are floated on a bath of water.  A steel sphere impacts the centre of the sheet at a speed $V$, deforming it out of plane; at the same time, a series of fine-scale wrinkles form, as illustrated in Figure~\ref{fig:snapshots}. However, this system also evolves dynamically: At first (panel a, $0.1\mathrm{~ms}$ after impact), very small-amplitude creases appear to propagate radially outward from the point of impact.  Next, an axisymmetric transverse wave forms and, immediately, fine wrinkles are observed to decorate the otherwise flat portion of the sheet  (panel b). Finally, as the transverse wave propagates, the wrinkles in this flat portion coarsen, increasing in wavelength (panels c and d). We do not address the (very) small-amplitude creases that appear immediately (within $0.1\mathrm{~ms}$) upon impact, but rather focus on the evolution of the axisymmetric wave and the exterior wrinkles.  Typical experimental parameter values are given in Table~\ref{table:params}, while a more detailed description of the experiments is given in~\cite{pnas}.

\begin{table}[ht]
\centering 
\begin{tabular}{l c c c} 
Parameter & Symbol & Range & Units \\ [0.5ex] 
\hline \hline \\[-2.00ex] 
 Sheet thickness & $\h$ & $1.5$--$5.3\times 10^{-7}$&$\mathrm{m}$ \\ 
 Sheet radius & $\rf$ & $5.51$--$17.72\times 10^{-3}$&$\mathrm{m}$\\
 Young's modulus & $E$ & $3.46\times 10^9 $&$\mathrm{N} \mathrm{m}^{-2}$\\
 Poisson's ratio & $\nu$ & 0.33 \\
 Sheet density & $\rhosheet$ & $1.05 \times 10^3$ & $\mathrm{kg}\mathrm{m}^{-3}$\\
 \hline\\[-3.00ex]
 Impactor radius & $\rs$ & $0.5$--$1.75\times 10^{-3}$&$\mathrm{m}$\\
 Impactor speed & $V$ & $0.78$--$2.85$&$\mathrm{m}\mathrm{s}^{-1}$\\
 Impactor density & $\rhosphere$ & $7720$&$\mathrm{kg}\mathrm{m}^{-3}$\\
 \hline\\[-3.00ex]
 Density of water & $\rho$ & $10^3 $&$\mathrm{kg} \mathrm{m}^{-3}$ \\
 Surface tension coefficient & $\gamma$ & $73\times 10^{-3} $&$\mathrm{N}\mathrm{m}^{-1}$\\
[1ex] 
\hline
\end{tabular}
\caption{Typical parameter values for the experimental study of impact--induced wrinkling studied presented in ref.~\cite{pnas}.  
  }
\label{table:params} 
\end{table}

In this paper, we develop a mathematical model to quantitatively describe these experiments.  In particular, our aim is to describe both the propagation of the transverse wave and the coarsening of the wrinkles, since both of these features are in contrast with what has been observed in static scenarios (see \S\ref{sec:static}): the typical horizontal extent of the curved region of a quasi-statically indented, wrinkled sheet is known to be a constant independent of the indentation depth. 
Similarly, the quasistatic arguments presented in~\S\ref{sec:static} would suggest that the number of wrinkles in the flat portion of the sheet should be determined by buoyancy and be independent of indentation depth, but in the dynamic experiment shown in Figure~\ref{fig:snapshots} we see that the number of wrinkles decreases with time (which corresponds to increasing indentation depth).

Despite the qualitative differences between previous static experiments on wrinkling and the current dynamic study, the phenomena outlined above are more familiar in hydrodynamic settings:  
the propagation of a transverse wave is at least qualitatively similar to the capillary waves that propagate when a stone is dropped into a pond, as suggested by~\cite{Vandenberghe2016} in a related problem. Similarly, the coarsening of the wrinkles is qualitatively similar to the coarsening of wrinkles in an elastic sheet on a thin viscous layer \cite{Vandeparre2010,Kodio2017}. In this paper we shall show that both of these phenomena are indeed governed by hydrodynamics, although we shall also see that the very fine-scale wrinkling of the sheets leads to a number of novel quantitative results, including new scaling laws.

\section{Mathematical model: dynamic point impact} \label{sec:intro}
\begin{figure}[ht]
\begin{tikzpicture}
\def\rad{0.2}
\def\top{1.5}
\def\mid{-0.5}
\def\bot{-1.5}

\draw (-\rad,\top) to (\rad,\top) to (\rad,\mid) to (0,\bot) to (-\rad,\mid) to (-\rad,\top);
\fill[gray!20] (-\rad,\top) to (\rad,\top) to (\rad,\mid) to (0,\bot) to (-\rad,\mid) to (-\rad,\top);

\draw[{<[scale=1.0]}-{>[scale=1.0]},dashed] (-1.5,-0.68-0.85) to (-1.5,0.0);
\node[left] at (-1.5,-0.75) {$Vt$};

\node[black] at (-5.0,-1.5)[rectangle,draw,fill=blue!20] {{Fluid}};
\node[black] at (-5.0,1.5)[rectangle,draw,fill=blue!20] {{Air}};
\draw[<-] (3.1,0.25) to (3.5,1.05) node[rectangle,draw,fill=blue!20] {Sheet};

\draw[->,dashed] (0,0) to (0.0,1.0) node[above]{$z$};
\draw[->,dashed] (0.0,0.0) to (1.0,0.0) node [above right] {$r$};
\draw[dotted] (-7,0.0) to (7,0.0);

\draw[->] (-3.0,0.0) to (-3.0,0.2);
\node[above] at (-3.0,0.2) {$w(r,\theta,t)$};

\begin{axis}[
    at = {(0,0)}, anchor = {origin},
    width = 0.7\textwidth,
    height = 0.205\textwidth,
    xmin = -3.5, xmax = 3.5,
    ymin = -1.02, ymax = 0.3,
    every axis plot/.append style={thick},
    axis line style={draw=none},
    ticks=none,
    ]
    \addplot[black] table {similarity-solution/intro/pointindenterprofileforintro.txt};
\end{axis}

\end{tikzpicture}
{\caption{Schematic diagram: a point impact on a floating elastic sheet, causes a vertical deflection of the sheet, $w(r,\theta,t)$, as well as fluid flow beneath the sheet.}
\label{fig:model0}}
\end{figure}
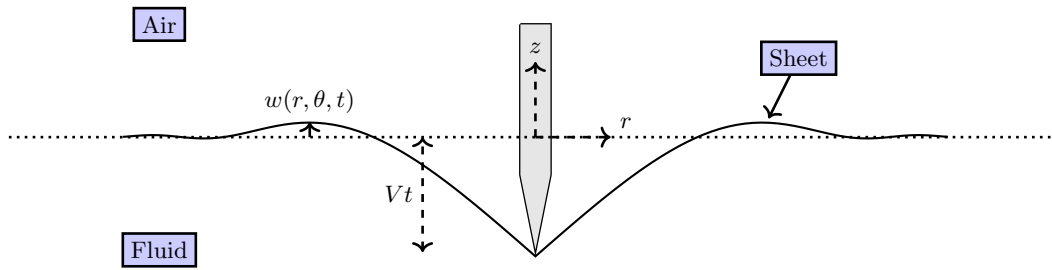

We consider a circular thin elastic sheet of radius $\rf$ and thickness $\h$, floating on the surface of an inviscid liquid of density $\rho$ and surface tension coefficient $\gamma$.  We use a cylindrical coordinate system centred on the centre of the sheet, $r = 0$, with $z = 0$ the initial vertical position of the undeformed sheet and with $\unitk$ the unit vector in the positive $z$ direction.  At time $t = 0$, a point impacts the centre of the sheet, moving downward at a fixed speed $V$ as illustrated in Figure~\ref{fig:model0}.

The elastic sheet deforms as a result of the impact; the components of this elastic deformation may be written $\boldsymbol{u} = (u_r, u_{\theta}, w)$, with each component varying in both space and time. (Note that throughout this paper we shall use $\boldsymbol{u}$ to denote the elastic displacements and $\boldsymbol{v}$ to denote fluid velocities.) 
The  components of the elastic displacement are coupled to one another:  vertical deflections of the sheet pull material circles radially inward, leading to a radial displacement $u_r<0$, which in turn causes the azimuthal (hoop) compression that leads to wrinkling. 

In the static indentation problem considered previously \cite{Vella2015PRL,Paulsen2016pnas} and summarized in \S\ref{sec:static}, the response of the elastic sheet to loading is determined by the hydrostatic pressure within the liquid, the tension of the bare liquid--air interface and the mechanical properties of the elastic sheet. This balance leads to the emergence of the horizontal lengthscale over which the sheet is curved $\lcurv\sim\rf^{1/3}\lc^{2/3}$, given in \eqref{eqn:lcurvStatic}. As already discussed, $\lcurv$ is independent of the depth of indentation but here the dynamic analog of $\lcurv$ grows in time, as shown in Figure~\ref{fig:snapshots}. We must therefore consider the motion of the fluid beneath the sheet due to impact, and the effect of this flow on the sheet.

\subsection{Fluid flow}

Since the impact occurs `fast' (the Reynolds number $\operatorname{Re} = \rho V \rs/\mu \sim 10^3$) and the fluid is initially stationary, we assume that the flow is a potential flow, i.e.~that the fluid velocity field $\boldsymbol{v}=\nabla\varphi$ for some velocity potential $\varphi(r,\theta,z,t)$.  (We note that the reduced Reynolds number $\operatorname{Re}^{*} = \rho V \lambda/\mu$, calculated using the wrinkle wavelength as a lengthscale, may be smaller, but for the smallest measured wavelengths $\lambda = O(100 \mathrm{\mu} \mathrm{m})$ in~\cite{pnas}, we still find $\operatorname{Re}^{*} \sim 10^2 \gg 1$.)
Assuming further that the fluid is incompressible, we have that
\begin{equation} \label{eq:laplace}
\nabla^2 \varphi = 0,
\end{equation} for $0\leq r<\infty$, $-\infty<z\leq w(r,\theta,t)$, where $w(r,\theta,t)$ denotes the shape of the deformed fluid interface (i.e.,~the sheet and liquid surface).  Far from the sphere, we expect the velocity potential to decay, so we choose $\varphi \to 0$ as $r^2+z^2\to\infty$, and at the origin we apply the symmetry condition that $\partial\varphi/\partial r=0$ at $r=0$. 

The fluid flow is forced by the kinematic boundary condition on the interface, $z=w(r,\theta,t)$, which reads
\begin{equation} \label{eq:dimkin}
\frac{\partial \varphi}{\partial z} = \frac{\mathrm{D}w}{\mathrm{D}t},
\end{equation} with $\mathrm{D}/\mathrm{D}t = \partial/\partial t + \boldsymbol{v} \cdot \nabla$ the convective derivative.   

For a point impactor, the surface is at $w = -Vt$ at $r = 0$ but free elsewhere, so we require a dynamic boundary condition to close the problem.  To write down this condition, we must examine the behaviour of the elastic sheet at the interface in more detail.

\subsection{Elastic deformation}  
The location of the free surface $z = w(r,\theta,t)$ is not known \emph{a priori}, and must be determined as part of the solution.  
The deformation of the elastic sheet at this interface is governed by the vertical force balance \cite{HowellAppliedSolidMechBook}
\begin{equation} \label{eq:dimvert}
\rhosheet h \frac{\partial^2 w}{\partial t^2}=p[r,\theta,w(r,\theta,t),t]- B \nabla^4 w + \nabla \cdot (\boldsymbol{\sigma} \cdot \nabla w ),
\end{equation} 
provided $|\nabla w| \ll 1$, i.e.,~for small slopes. 
Here $p(r,\theta,z,t)$ is the pressure within the liquid (measured relative to atmospheric pressure above the sheet), $\rhosheet$ and $B = E \h^3/[12(1-\nu^2)]$ are the density and bending stiffness of the elastic sheet, respectively (with $E$ and $\nu$ the Young's modulus and Poisson's ratio, respectively), and $\boldsymbol{\sigma}$ is the two-dimensional (thickness-integrated) stress  in the sheet.  We will later assume the sheet radius is large compared with the wavefront position, so we will not concern ourselves with the deformation of the liquid--air interface beyond the sheet.  

To progress further, we need to determine the unknown pressure $p(r,\theta,z,t)$ in the liquid just beneath the elastic sheet. If the fluid velocity potential is known, then $p$ may be determined from the Bernoulli equation
\begin{equation} \label{eq:bernoulli2}
\rhol \left( \frac{\partial \varphi}{\partial t} + \frac{1}{2} |\nabla \varphi|^2 \right) + p + \rhol g z = c(t),
\end{equation} with $c(t)$ to be determined. We can then eliminate the pressure from \eqref{eq:dimvert} by evaluating \eqref{eq:bernoulli2} at the free surface to arrive at
\begin{equation} \label{eq:dimdyn}
\rhol \left( \frac{\partial \varphi}{\partial t} + \frac{1}{2} |\nabla \varphi|^2 \right) + \rhosheet h \frac{\partial^2 w}{\partial t^2} + B \nabla^4 w - \nabla \cdot \left( \boldsymbol{\sigma} \cdot \nabla w \right) + \rhol g w = 0,
\end{equation}
where $c(t) = 0$ has been chosen by an appropriate choice of $\varphi$.

The final piece of the puzzle is to determine the stress distribution within the solid sheet, $\boldsymbol{\sigma}$, governed by  
\begin{equation} \label{eq:dimhorz}
\rhosheet \h \frac{\partial^2 \boldsymbol{u}}{\partial t^2} = \nabla \cdot \boldsymbol{\sigma},
\end{equation} 
with the radial tension in the sheet equal to the surface tension at the adjacent liquid--air interface, i.e.\ 
\begin{equation} \label{eq:edgebc}
\srrf(r=\rf) = \gamma.
\end{equation} 

In addition to satisfying \eqref{eq:dimhorz}, the in-plane stresses within the sheet are related to the displacement field via the strain tensor $\boldsymbol{\epsilon}$ and Hooke's law \cite{HowellAppliedSolidMechBook}:
\begin{subequations} \label{eq:dimstressstrain}
\begin{equation} \label{eq:dimstrainrr}
\frac{1}{E\h} \left( \srrf - \nu \sttf \right)=\epsrr = \frac{\partial u_r}{\partial r} + \frac{1}{2} \left( \frac{\partial w}{\partial r} \right)^2,
\end{equation}
\begin{equation} \label{eq:dimstraintt}
\frac{1}{E\h} \left( \sttf - \nu \srrf \right)=\epstt = \frac{u_r}{r} + \frac{1}{r} \frac{\partial u_{\theta}}{\partial \theta} + \frac{1}{2} \left( \frac{1}{r} \frac{\partial w}{\partial \theta} \right)^2,
\end{equation}
\begin{equation}
 \frac{1 + \nu}{E\h} \srtf=\epsrt = \frac{1}{2} \left( \frac{1}{r} \frac{\partial u_r}{\partial \theta} + \frac{\partial u_{\theta}}{\partial r} - \frac{u_{\theta}}{r}\right) + \frac{1}{2} \left(\frac{\partial w}{\partial r} \right) \left( \frac{1}{r} \frac{\partial w}{\partial \theta} \right).
\end{equation}
\end{subequations} 
These equations may be used to relate the horizontal and vertical displacement components and ultimately determine how much length must be absorbed by wrinkling.  For sufficiently stiff materials that $(\srrf,\sttf)/(Eh)\to0$, eqn \eqref{eq:dimstraintt} shows the amplitude of the wrinkles is set by ensuring they absorb all of the excess length generated by pulling material circles to a smaller radial position, $u_r<0$.

\subsection{Non-dimensionalization} \label{sec:nondim-action}
To non-dimensionalize, we need to identify an appropriate lengthscale.  The scaling argument of ref.~\cite{pnas} was based on a radial stress in wrinkled sheets $\sigma_{rr} = \gamma \rf/r$, and showed that the radial wavefront of the transverse wave $r_m \sim \left(\gamma \rf t^2/ \rho\right)^{1/4}$. Comparing fluid inertia and radial tension terms in~\eqref{eq:dimdyn}, with $\partial\varphi/\partial t \sim V^2$ and $\partial^2/\partial r^2 \sim 1/r_m^2$, then suggests a horizontal lengthscale
\begin{equation}
\lstar = \sqrt{\frac{\gamma \rf}{\rhol \V^2}}.
\end{equation}
We shall see below that this lengthscale is associated with the constant radius of curvature of the impact `center'.  

Based on these observations, we introduce dimensionless variables
\begin{subequations} \label{eq:nondim}
\begin{equation}
 (\hat{r}, \hat{z}, \hat{w} )=(r,z,w)/\lstar, \quad \hat{t} = \frac{\V}{\lstar} t , \quad \hat{\varphi}= \varphi/(\V \lstar) ,
\end{equation}
\begin{equation}
\hat{u}_i =u_i/\lstar , \quad \hat{\sigma}_{ij} =\sigma_{ij}/\sigmastar,
\end{equation}
\end{subequations}
where $\sigmastar = \gamma \rf/\lstar$.
We can immediately see that the effects of sheet inertia and gravity in~\eqref{eq:dimdyn}--\eqref{eq:dimhorz} are negligible over these scales: the inertia of the sheet relative to the fluid inertia in~\eqref{eq:dimdyn} and relative to the stress term in~\eqref{eq:dimhorz} is approximately $\rhosheet \h/\rhol \lstar = O(10^{-3})$, and the importance of gravity relative to dynamic pressure is given by the square of the inverse Froude number $g \lstar/V^2 = O(10^{-3})$--$O(10^{-2})$.

Finally, for very thin sheets the bending stress term $B \nabla^4 w$ in~\eqref{eq:dimdyn} will be small compared to the other terms.  However, it is known that the limit of very small bending stiffness is singular \cite{Davidovitch2011}: this term is therefore expected to become important over very short lengthscales, as is the case, for example, with very fine scale wrinkles.  Therefore, we turn our attention to the wrinkle lengthscale before completing our non-dimensionalization.

\subsection{The wrinkling lengthscale} \label{sec:dominantbalance}

Previous analyses of highly wrinkled elastic sheets~\cite{Davidovitch2011,Vella2015PRL, Taffetani2017} consider the vertical deflection of the sheet to be composed of a leading-order axisymmetric displacement, $\bar{w}(r)$, decorated by small amplitude, short wavelength wrinkles with a well-defined wavenumber $m\gg1$, i.e.,~they use an ansatz
\beq
w(r,\theta)=\bar{w}(r)+m^{-1}f(r)\cos m\theta
\eeq where the function $f(r)$ is assumed to be an $O(1)$ quantity and the appearance of $1/m \ll 1$ encodes the expectation that the wrinkles have a short wavelength and small amplitude, but collectively absorb the $O(1)$ amount of excess length that is introduced via the leading-order mean shape $\bar{w}(r)$.  
As with previous static studies \cite{Paulsen2016pnas,Taffetani2017}, we shall assume that spatial variations in the wrinkle wavenumber $m$ occur over a lengthscale that is large compared to the wrinkle wavelength. From a perturbation methods point of view, a new complication of the current problem is that this wrinkle number may vary with time, so using $m^{-1}$ as our asymptotically small parameter would introduce unnecessary complications.  Instead, we define a small parameter $\spar \ll 1$ such that the wavelength $\lambda = O(\spar \lstar)$, i.e.,~$m = O(\spar^{-1})$.  Comparing the  $u_r$ and $w$ terms in the expression for azimuthal strain~\eqref{eq:dimstraintt}, we observe that when a radial contraction of $O(\lstar)$ induces wrinkles of wavelength $O( \spar\lstar)$, those wrinkles must have height $O(\spar\lstar)$. This tells us that both the wavelength and the amplitude of the wrinkles are smaller than the global lengthscale by the unknown factor $\spar \ll 1$.  {We note that, although the amplitude and wavelength of the wrinkles are of the same order of magnitude in comparison to $\lstar$, the slope of the wrinkles is small throughout their evolution, as we shall verify in due course.}

To progress further, we must determine the parameter $\spar$ (the mismatch between the wrinkling and global lengthscales) in terms of physical parameters.  In static problems, $\spar$ is determined by minimizing the energy in the system~\cite{Taffetani2017}; since the system under consideration here is dynamic, Hamilton's principle of minimal action may be used to determine the appropriate relationship.  For simplicity, however, we use here a dominant balance argument.    We consider the various terms of~\eqref{eq:dimdyn}, neglecting the inertia of the sheet itself and the effect of gravity based on the arguments in \S\ref{sec:nondim-action}.  Since we do not yet know the timescale or the size of the velocity potential associated with wrinkling, we first seek a balance between the bending, compression and tension terms associated with wrinkling in~\eqref{eq:dimdyn} to find
\begin{equation} \label{eq:LargeMBalance1}
\frac{B m^4(\lstar/m)}{\lstar^4}\sim\frac{\sigma_\ast^{\theta\theta} m^2\lstar/m}{\lstar^2}\sim \frac{\sigmastar \lstar/m}{\lstar^2},
\end{equation}
where $\sigma_{\ast}^{\theta\theta}$ is the characteristic size of the hoop stress.

Examining the balance between the second and third terms in~\eqref{eq:LargeMBalance1}, we immediately see that $\sigma_{\ast}^{\theta\theta} \sim m^{-2} \sigmastar\ll\sigmastar$; when used to estimate the size of terms in the equation of in-plane equilibrium~\eqref{eq:dimhorz}, we find that $\partial(r\sigma_{rr})/\partial r\approx0$ and hence, using the boundary condition~\eqref{eq:edgebc} verify our choice $\sigmastar \sim \gamma \rf/\lstar$.   Balancing with the bending term then gives

\beq
m\sim \left(\frac{\gamma\rf\lstar}{B}\right)^{1/4}
\label{eqn:mscale}
\eeq 
and hence we choose
\beq
\spar=\left(\frac{B}{\gamma\rf\lstar}\right)^{1/4}\ll1.
\label{eqn:deltadefn}
\eeq
We note that the scaling for $m$ in \eqref{eqn:mscale} is similar to the typical static scaling $m\sim (TR^2/B)^{1/4}$ for {static} wrinkles formed by a tension $T$ over some lengthscale $R$~\cite{Davidovitch2011}; our definition of $\spar$ is essentially the fourth root of the inverse bendability defined by Davidovitch \emph{et al.} \cite{Davidovitch2011}. 
We now rescale $\theta$ and introduce a wrinkling lengthscale and timescale according to
\begin{equation} \label{eq:rescalings}
\theta = \spar \btheta, \quad z = \spar \lstar \bz. 
\end{equation} 

Finally, we turn our attention to the momentum term on the left-hand side of~\eqref{eq:dimdyn}, and ask when fluid inertia will affect wrinkling behaviour. In the flat portion of the sheet, we anticipate that the {advective} term {which} is quadratic in velocity will be small, and the effect of inertia will be limited to the acceleration term, $ \rho \partial \phi_w/\partial t$, where $\phi_w$ is the velocity potential associated with wrinkling.  Seeking a balance both in the kinematic boundary condition~\eqref{eq:dimkin} (with vertical scale $\bz=O(1)$) and between the inertia term in~\eqref{eq:dimdyn} and the other terms highlighted in~\eqref{eq:LargeMBalance1}, we observe that this is only possible on a shorter timescale $t = O(\spar^{1/2})$.

We emphasize that wrinkling takes place on a timescale that is a factor $\spar^{-1/2}$ faster than that used in the non-dimensionalization (\ref{eq:nondim}a).  This suggests that as $\spar \rightarrow 0$ {the} wrinkles {in the flat portion of the sheet} should appear quasistatic on the timescale associated with impact and the propagation of the axisymmetric wave.  However, for the experiments reported in~\cite{pnas}, $\spar^{1/2} \approx 0.3$;  these timescales are therefore not well-separated and we expect the wrinkles to continue to coarsen at the same time as impact proceeds --- as is observed experimentally. Nevertheless, considering the asymptotic limit $\spar \rightarrow 0$ allows us to derive simplified models approximating the evolution of both the axisymmetric wave and the wrinkles.

\subsection{Dimensionless governing equations}
We treat the two vertical variables $z$ and $\bz$ (associated with global deformation and wrinkling, respectively) as independent, so that, using the chain rule, spatial derivatives in the vertial direction are rewritten $\spar^{-1} \partial/\partial \bz + \partial/\partial z$.  
We now substitute the dimensionless variables~\eqref{eq:nondim} and~\eqref{eq:rescalings} into the governing equations~\eqref{eq:laplace}--\eqref{eq:dimdyn}.  Taking care to include the derivatives associated with all length and timescales, the Laplace equation~\eqref{eq:laplace} for the velocity potential is transformed to
\begin{equation} \label{eq:laplace2}
\left[ \frac{1}{r} \frac{\partial}{\partial r} \left( r \frac{\partial \varphi}{\partial r} \right) + \frac{\partial^2 \varphi}{\partial z^2} \right] + 2 \spar^{-1} \frac{\partial^2 \varphi}{\partial z \partial \bz} + \spar^{-2} \left[ \frac{1}{r^2} \frac{\partial^2 \varphi}{\partial \btheta^2} + \frac{\partial ^2 \varphi}{\partial \bz^2} \right] = 0,
\end{equation} 
where here, and henceforth, we have dropped hats from dimensionless variables. The kinematic condition~\eqref{eq:dimkin} is transformed to
\begin{equation} \label{eq:kinematic}
\spar^{-1} \frac{\partial \varphi}{\partial \bz} + \frac{\partial \varphi}{\partial z} = \frac{\partial w}{\partial t} + \frac{\partial w}{\partial r} + \frac{\partial \varphi}{\partial r} \frac{\partial w}{\partial r} + \spar^{-2} \left(\frac{1}{r} \frac{\partial \varphi}{\partial \btheta} \right) \left( \frac{1}{r} \frac{\partial w}{\partial \btheta} \right).
\end{equation}

Based on the scaling arguments in \S\ref{sec:nondim-action}, we neglect the inertia of the sheet so that the Bernoulli equation~\eqref{eq:dimdyn} and horizontal force balance~\eqref{eq:dimhorz} are transformed to
\begin{multline} \label{eq:vertforce}
\frac{\partial \varphi}{\partial t} + \frac{1}{2} \left[ \left( \frac{\partial \varphi}{\partial r} \right)^2 + \left( \frac{\partial \varphi}{\partial z} \right)^2 \right] + \spar^{-1} \frac{\partial \varphi}{\partial z} \frac{\partial \varphi}{\partial \bz} + \frac{1}{2} \spar^{-2} \left[ \left( \frac{1}{r} \frac{\partial \varphi}{\partial \btheta} \right)^2 + \left( \frac{\partial \varphi}{\partial \bz} \right)^2 \right].
\\ + \left[ \spar^2 \frac{1}{r} \frac{\partial}{\partial r} \left( r \frac{\partial}{\partial r} \right) + \frac{1}{r^2} \frac{\partial^2}{\partial \btheta^2} \right]^2 w
 - \srrf \frac{\partial^2 w}{\partial r^2} - \sttf \left( \spar^{-2} \frac{1}{r^2} \frac{\partial^2 w}{\partial \btheta^2} + \frac{1}{r} \frac{\partial w}{\partial r} \right) = 0,
\end{multline}
and
\begin{subequations} \label{eq:horzforce}
\begin{equation} \label{eq:horzforcer}
\frac{1}{r} \frac{\partial}{\partial r} \left( r \srrf \right) + \spar^{-1} \frac{1}{r} \frac{\partial \srtf}{\partial \btheta} - \frac{\sttf}{r} = 0,
\end{equation}
\begin{equation} \label{eq:horzforcet}
\frac{1}{r} \frac{\partial}{\partial r} \left( r \srtf \right) + \spar^{-1} \frac{1}{r} \frac{\partial \sttf}{\partial \btheta} + \frac{\srtf}{r} = 0
\end{equation}
\end{subequations}
respectively, where we have used~\eqref{eq:horzforce} to rewrite the last two terms in~\eqref{eq:vertforce}.  The horizontal force balance~\eqref{eq:horzforce} must be solved subject to the stress conditions
\begin{equation} \label{eq:edgestress}
\srrf = \frac{1}{\rf/\lstar}, \quad \srtf = 0 \quad \text{at} \quad r = \frac{\rf}{\lstar}.
\end{equation}
The stress--strain relationships~\eqref{eq:dimstressstrain} are now given by
\begin{subequations} \label{eq:stressstrain}
\begin{equation}\label{eq:stressstrainrr}
\frac{\sigmastar}{E\h} \left( \srrf - \nu \sttf \right) = \epsrr = \frac{\partial u_r}{\partial r} + \frac{1}{2} \left( \frac{\partial w}{\partial r} \right)^2,
\end{equation}
\begin{equation} \label{eq:stressstraintt}
\frac{\sigmastar}{E\h} \left( \sttf - \nu \srrf \right) = \epstt = \frac{u_r}{r} + \spar^{-1} \frac{1}{r} \frac{\partial u_{\theta}}{\partial \btheta} + \frac{1}{2} \spar^{-2} \left( \frac{1}{r} \frac{\partial w}{\partial \btheta} \right)^2,
\end{equation}
\begin{equation} \label{eq:stressstrainrt}
 \frac{\sigmastar}{E\h} (1 + \nu) \srtf = \epsrt = \frac{1}{2} \left(  \spar^{-1} \frac{1}{r} \frac{\partial u_r}{\partial \btheta} + \frac{\partial u_{\theta}}{\partial r} - \frac{u_{\theta}}{r}\right) 
+ \frac{1}{2} \spar^{-1} \left(\frac{\partial w}{\partial r} \right) \left( \frac{1}{r} \frac{\partial w}{\partial \btheta} \right).
\end{equation}
\end{subequations}
The conditions at the origin are
\begin{equation} \label{eq:origin}
w = -t, \quad \frac{\partial \varphi}{\partial r} = 0 \quad \text{at} \quad r = 0,
\end{equation}
and the far-field conditions are
\begin{subequations} \label{eq:farfieldall}
\begin{equation} \label{eq:farfieldphi}
\varphi \rightarrow 0 \quad \text{as} \quad r^2 + z^2 \rightarrow \infty,
\end{equation}
\begin{equation} \label{eq:farfieldwrinkle}
\frac{\partial \varphi}{\partial \btheta}, \hspace{1.5mm}  \frac{\partial \varphi}{\partial \bz} \rightarrow 0 \quad \text{as} \quad \bz \rightarrow -\infty,
\end{equation}
\begin{equation}
w \rightarrow 0 \quad \text{as} \quad r \rightarrow \infty,
\end{equation}
\end{subequations}
where we have required that the perturbation to the fluid flow arising from wrinkling be localized near the interface.

\subsection{Oscillatory and non-oscillatory components}
Throughout our analysis we will find it beneficial to split variables into oscillatory and non-oscillatory components, i.e. we write
\begin{equation}
f = \bar{f} + \R{f}, \quad \text{with} \quad \frac{1}{2 (\pi/\spar) } \int_{-\pi/\spar}^{\pi/\spar} f\ \mathrm{d} \btheta = \bar{f}.
\end{equation}
We will also split the governing equations into averaged and oscillatory components, when it is useful to do so.

\subsection{High-bendability limit} \label{sec:highbendabilitylimit}
We now consider the limit $\spar \rightarrow 0$, which corresponds to the sheet having a low bending stiffness, or being highly `bendable'~\cite{Davidovitch2011}.  Based on the static case~\cite{Vella2015PRL}, we anticipate that the sheet experiences azimuthal compression and wrinkles everywhere except in some small region near the impactor, where the absolute azimuthal stress becomes positive again.
We will initially assume that this `tensile core' is small, and verify this in \S\ref{sec:tensilecore}.  
We begin by focusing on the sheet dynamics on the $O(1)$ timescale associated with the axisymmetric deformation of the sheet.  Following from our non-dimensionalization~\eqref{eq:nondim}, the variables associated with this motion should appear at $O(1)$.  Recalling our argument in~\S\ref{sec:nondim-action}, we expect that the vertical deformation associated with wrinkling (and hence depending on $\btheta$) should appear at $O(\spar)$.  We therefore propose an asymptotic expansion of the form
\begin{equation}
w = \bw{0}(r,t) + \spar \w{1}(r,\btheta,t) + \cdots,
\end{equation}
and similarly for other dependent variables.   Focusing on the velocity potential, we make two notes: (i) since the ``short'' vertical variable $\bz$ is associated with wrinkling and $\btheta$-dependence, we assume that dependence on $\bz$ appears at the same order as $\btheta$-dependence, and (ii) a velocity potential associated with the wrinkles must be induced through the kinematic boundary condition~\eqref{eq:kinematic0}, and therefore is of size $\spar^2$ (recalling that, for the moment, we consider only the $t=O(1)$ timescale, rather than the faster, wrinkling, timescale.  Based on these observations, we write
\begin{equation}
\varphi = \bp{0}(r,z,t) + \spar \bp{1}(r,z,t) + \spar^2 \p{2}(r,\btheta,z,\bz,t) + \cdots.
\end{equation}

\section{Axisymmetric elastocapillary wave} \label{sec:elastocapillarywave}

\subsection{Asymptotic analysis} \label{sec:elastocapillarywave-1}
We now derive a simplified model governing the leading-order axisymmetric out-of-plane deflection caused by a point impactor; we do this by substituting our asymptotic expansion into the dimensionless governing equations~\eqref{eq:laplace2}--\eqref{eq:farfieldall}, focusing first on the \emph{azimuthally averaged} velocity potential, and evaluating the Laplace equation~\eqref{eq:laplace2} order-by-order.   
At leading order, we find 
\begin{equation} \label{eq:laplace3} 
\frac{1}{r} \frac{\partial}{\partial r} \left( r \frac{\partial \bp{0}}{\partial r} \right) + \frac{\partial^2 \bp{0}}{\partial z^2} = 0,
\end{equation}
with 
\begin{subequations} \label{eq:lo-bcs1}
\begin{equation} \label{eq:lo-origin}
\frac{\partial \bp{0}}{\partial r} = 0 \quad \text{on} \quad r = 0,
\end{equation}
\begin{equation}\label{eq:lo-far-phi}
\bp{0} \rightarrow 0 \quad \text{as} \quad r^2 + z^2 \rightarrow \infty.  
\end{equation}
\end{subequations}
At the free surface $z \approx \bw{0}(r,t)$, the fluid velocity and the motion of the elastic sheet are coupled through the kinematic condition~\eqref{eq:kinematic}
\begin{equation}\label{eq:kinematic-lo-a}
\frac{\partial \bp{0}}{\partial z} = \frac{\partial \bw{0}}{\partial t} + \frac{\partial \bp{0}}{\partial r} \frac{\partial \bw{0}}{\partial r}. 
\end{equation}
At $r=0$, the deformation of the elastic sheet is simply
\begin{equation} \label{eq:defcontact0}
\bw{0} = - t , \quad \bur{0} = \but{0} = 0.
\end{equation}
For $r > 0$, the profile of the interface is not known \emph{a priori}, and so the kinematic condition
must be supplemented by the force balance and stress-strain relations~\eqref{eq:vertforce}--\eqref{eq:stressstrain}.  We now examine the forces outside the contact region order by order.  

We first observe that if there is wrinkling with $\partial^2 \w{1}/\partial \btheta^2 \neq 0$, then the vertical force balance~\eqref{eq:vertforce} dictates that $\sigma_{\theta\theta} = O(\spar)$ at most, so the hoop stress collapses, as is known in the limit of high bendability for static deformations~\cite{Davidovitch2011,Vella2015PRL,Paulsen2016pnas}.  Examining the horizontal force balances and using the boundary conditions~\eqref{eq:edgestress}, we can now deduce that
\begin{equation}
\sigma_{rr} = \frac{1}{r} + O(\spar), \quad \sigma_{r\theta} = \spar^2 \hat{\sigma}_{r\theta}^{(2)} + O(\spar^3), \quad  \frac{\partial \sigma_{\theta\theta}}{\partial \btheta} = O(\spar^3).
\end{equation}
Similarly, evaluating the vertical force balance~\eqref{eq:vertforce} at $O(1)$, separating out axisymmetric and oscillatory components,  and using the fact that $\sigma_{\theta\theta}$ is independent of $\btheta$ before $O(\spar^2)$, we find that the leading-order variables satisfy
\begin{equation} \label{eq:dynlo-a}
\frac{\partial \bp{0}}{\partial t} + \frac{1}{2} \left[ \left( \frac{\partial \bp{0}}{\partial r} \right)^2 + \left( \frac{\partial \bp{0}}{\partial z} \right)^2 \right] %
= \frac{1}{r} \frac{\partial^2 \bw{0}}{\partial r^2},
\end{equation} 
together with the result that $\sttb{1} = 0$, 
so that the leading-order hoop stress is $\spar^2 \sttb{2}$, and the far-field condition 
\begin{equation} \label{eq:lo-far-w}
\bw{0} \rightarrow 0 \quad \text{as} \quad r \rightarrow \infty.
\end{equation}  
Physically, the result \eqref{eq:dynlo-a} may be interpreted as the hydrodynamic pressure in the liquid (the left-hand side of \eqref{eq:dynlo-a}) being balanced by the pressure jump across the sheet ($\srrf\kappa_{rr}$), with $\srrf$ being the stress in the wrinkled sheet, which is qualitatively altered by wrinkling.  

We now have a complete description of the leading-order stresses, out-of-plane deformation $\bw{0}$, and velocity potential $\bp{0}$.  We can also write down the leading-order in-plane deformation at this stage: examining the stress--strain relations~\eqref{eq:stressstrain} we see that $\but{0} = 0$, while the radial deformation is given by
\begin{equation} \label{eq:radialdef}
\frac{\mathrm{d} \bur{0}}{\mathrm{d} r} = - \frac{1}{2} \left( \frac{\mathrm{d} \bw{0}}{\mathrm{d} r} \right)^2 + \frac{\sigma_{rr}^{\star} }{E\h} \frac{1}{r}, \quad \bur{0}(r = 0) = 0.
\end{equation}

By considering the high-bendability limit, $\spar \rightarrow 0$, we have determined that the leading-order sheet profile, together with the stress state within it, are axisymmetric; wrinkles are smaller in amplitude, with a wavelength $O(\spar \lstar)$ dictated by the bending stiffness in the sheet. We shall now solve the leading-order problem for the velocity potential $\bp{0}(r,z,t)$ and the mean vertical surface deformation $\bw{0}(r,t)$. We will find that this leading-order behaviour admits a similarity solution, which describes the propagation of the ripple that is observed experimentally \cite{pnas}.

\subsection{Similarity solution} \label{sec:simsol}
In dimensionless terms, the sheet edge is located at $r = \rf/\lstar$, which lies in the range $9.77\lesssim\rf/\lstar\lesssim44$ in the experiments of~\cite{pnas}.  At this point in our analysis, we further simplify our model by considering the limit $\rf/\lstar \rightarrow \infty$, so that the elastic sheet has infinite extent.  The leading-order, axisymmetric motion is described by a velocity potential $\bp{0}$ and vertical surface deformation $\bw{0}$, governed by the bulk equation~\eqref{eq:laplace3} together with boundary conditions~\eqref{eq:lo-bcs1}--\eqref{eq:lo-far-w}.  We shall focus on the early-time behaviour of the sheet, when the mean deflection $\bw{0}(r,t)$ is small enough to neglect hydrostatic pressure, and time-derivatives dominate advection. (We note that, since we have assumed that the sheet is completely wrinkled, our analysis only concerns intermediate times in a sense that we quantify in \S\ref{sec:tensilecore}.)

The scaling estimates of \cite{pnas} suggests that an evolving horizontal lengthscale $\propto t^{1/2}$ should be expected here, in contrast with the $r \sim t^{2/3}$ scaling expected for  impacts dominated by a constant tension~\cite{KellerMiksis1983,VellaMetcalfe2007,OKiely2015-meniscus,Vandenberghe2016}.  
Based on these observations, we seek a self-similar early-time solution 
\begin{equation} \label{eq:selfsim}
\bw{0}(r,t) \approx t \Hsim(\xi) , \quad \bp{0}(r,z,t) \approx t^{1/2}\f(\xi,\zeta), 
\end{equation}
with similarity variables
\begin{equation}
(\xi, \zeta) = \left( \frac{r}{\sqrt{t}}, \frac{z}{\sqrt{t}} \right).
\end{equation}

Carrying out a change of variables, the Laplace equation~\eqref{eq:laplace3} yields
\begin{equation} \label{eq:laplace4}
\frac{1}{\xi} \frac{\partial}{\partial \xi} \left( \xi \frac{\partial \f}{\partial \xi} \right) + \frac{\partial^2 \f}{\partial \zeta^2} = 0.
\end{equation}
The boundary conditions at the free surface should be applied at $\zeta = w/\sqrt{t} \approx \sqrt{t} \Hsim(\xi)$.  For $t \ll 1$, this corresponds to the fixed boundary $\zeta = 0$, which reflects the fact that for $t \ll 1$ the free surface is only slightly deformed.
The kinematic free-boundary condition~\eqref{eq:kinematic-lo-a} is transformed to
\begin{equation} \label{eq:kinematic0}
\frac{\partial \f}{\partial \zeta} = \Hsim - \tfrac{1}{2} \xi \frac{\mathrm{d} \Hsim}{\mathrm{d} \xi}
\end{equation}
and the dynamic boundary condition~\eqref{eq:dynlo-a} is transformed to
\begin{equation} \label{eq:simdyn}
\frac{1}{2} \left( \f - \xi \frac{\partial \f}{\partial \xi} \right) - \frac{1}{\xi} \frac{\mathrm{d}^2 \Hsim}{\mathrm{d} \xi^2} = 0,
\end{equation}
both on $\zeta = 0$.   We note that the non-linear terms in both boundary conditions are neglected in the limit $t \ll 1$.  Finally, the conditions at $r = 0$~\eqref{eq:lo-origin} and in the far field~\eqref{eq:lo-far-phi}, \eqref{eq:lo-far-w} may be written
\begin{subequations} \label{eq:origin0}
\begin{equation}
\Hsim = -1,
\end{equation}
\begin{equation}
\frac{\partial \f}{\partial \zeta} = -1,
\quad
\frac{\partial \f}{\partial \xi} = 0
\end{equation}
\end{subequations}
at $\xi = 0$, and 
\begin{subequations} \label{eq:farfield}
\begin{equation}
\Hsim \rightarrow 0 \quad \text{as} \quad \xi \rightarrow \infty,
\end{equation}
\begin{equation}
\f \rightarrow 0 \quad \text{as} \quad \zeta^2 + \xi^2 \rightarrow \infty.
\end{equation}
\end{subequations} 
We note from~\eqref{eq:simdyn} that the radial curvature $\partial^2 \w{0}/\partial r^2 \approx \mathrm{d}^2 \Hsim/d\mathrm \xi^2$ is independent of time.  We deduce that, in dimensional terms, the curvature $\sim 1/\lstar$; this reaffirms that the lengthscale $\lstar$ is the most natural choice for our non-dimensionalization.

The equations~\eqref{eq:laplace4}--\eqref{eq:farfield} were solved numerically using second-order finite differences.  We note that there are no free parameters, so the problem only needs to be solved numerically once.  The free surface is illustrated (together with numerical results for a finite-size impactor, see \S\ref{sec:finitealpha}) in Figure~\ref{fig:similarity}.  We emphasize that this self-similar solution is different to the one described by Philippi et al.~\cite{Philippi2016}: our solution is valid in the limit of a point impactor on a floating elastic sheet where surface tension dominates, while the solution of~\cite{Philippi2016} is valid in the limit of a large impactor at an interface where surface tension is negligible and fluid flow dominates the governing equations.

\subsection{Radial retraction} \label{sec:retraction}
Having determined the mean deflection profile of the sheet induced by impact, we now turn our attention to the corresponding radial displacement of material inwards, since this is what drives the observed wrinkling. The amount of retraction, $\ur{0}$, is determined by  the  axisymmetric similarity solution described above (to leading-order in the high-bendability limit $\spar \rightarrow 0$).  Comparing terms in~\eqref{eq:radialdef}, we note that the tension term is smaller than the vertical deformation terms by a factor $\rho V^2 \lstar/Eh = O(10^{-3})$, and may be neglected; we therefore find the radial retraction at early time
\begin{equation} \label{eq:retraction}
\bur{0}(\xi) \approx - \frac{1}{2} t^{3/2} \int_{0}^{\xi} \left( \frac{\mathrm{d} \Hsim}{\mathrm{d} \xi'} \right)^2\ \mathrm{d} \xi'.
\end{equation}
Taking the limit $\xi \rightarrow \infty$, we find that the retraction of the edge  $-\ur{0}(\infty)\approx0.41t^{3/2}$.  
This radial retraction in turn drives the formation of wrinkles in the sheet, which we investigate in \S\ref{sec:wrinkles}. 

Finally, we note that the radial retraction of the sheet over the liquid surface will induce a Blasius-like viscous boundary layer in the underlying fluid, and that this may in turn be expected to exert a shear on the sheet, slowing down the retraction.  To assess the importance of this viscous boundary layer, we consider a flat plate moving at a variable speed $v_r=\partial u_r/\partial t \sim t^{1/2}$ across the fluid surface.  Including viscous terms in the horizontal force balance, we find that they balance with the inertial term $\sim \partial v_r/\partial t$ over a dimensionless depth $z \sim \sqrt{t/\operatorname{Re}}$.  We estimate the effect of the shear $\sim \partial v_r/\partial z$ on the radial stress in the sheet, and find it induces a relative change of size $1/\sqrt{\operatorname{Re}} \ll 1$, independent of time and radial position.  We deduce that viscous shear should not have a significant effect on the retraction of the sheet.

\subsection{Conditions for a fully wrinkled sheet} \label{sec:tensilecore}
In our analysis of the elastic sheet deformation caused by impact, we have considered a scenario in which the entire sheet is covered with radial wrinkles.  This simplifies matters considerably, since even the static case is very involved {when a partially wrinkled sheet is considered}~\cite{Vella2018}.  We now provide scaling estimates of when this assumption, and hence our analysis, is valid.  Before impact, the sheet is flat with a uniform, isotropic tension, and disturbances can be expected to propagate according to the usual capillary wave relationship (in dimensionless terms) $r_m \sim \left(\lstar/\rf \right)^{1/3} t^{2/3}$.  When the impact depth is sufficiently small, the change in the stress in the sheet is also small, and the sheet remains under approximately isotropic tension with $\sigma_{rr} \approx \sigma_{\theta\theta} \approx \gamma/\sigmastar$.   The vertical deformation induces a change in this stress, so writing $\sigma_{rr} = \gamma/\sigmastar + \bar{\sigma}_{rr}$ and considering the stress--strain relationship~\eqref{eq:stressstrainrr} we find $\bar{\sigma}_{rr} \sim (Eh/\sigmastar) (\partial w/\partial r)^2$.  We expect that the sheet will become wrinkled when the stress induced by impact is comparable to the pre-stress due to surface tension.  Using the scalings  $r_m \sim (\lstar t^2/\rf)^{1/3}$, $w \sim t$ suggests that this wrinkle initiation occurs at a dimensionless time 
\begin{equation}
t_{\text{wrinkle}} \sim \left( \frac{\gamma}{E h} \right)^{3/2} \frac{\lstar}{\rf}.
\end{equation}
For the parameters in Table~\ref{table:params}, $t_{\text{wrinkle}} = O(10^{-7})$, corresponding to dimensional times on the order of $10^{-10}\,\mathrm{s}$, significantly earlier than any experimental observation in~\cite{pnas}.

Wrinkling is synonymous with an anisotropic and inhomogenous stress field in the sheet: the sheet does not wrinkle everywhere at the same time.  Based on the static case studied in detail in~\cite{Vella2018} we consider a wrinkled annulus $\lin < r < \lout$ and determine how the wrinkled annulus grows as time (and impact depth) increases.  In $r > \lout$, the sheet is wrinkled and the stresses are given by the Lam\'{e} solution
\begin{equation}
\sigma_{rr/\theta\theta} = \frac{\rf}{\lstar} \frac{1}{ \left(\rf/\lstar\right)^2 + \lout^2} \left[ 1 \pm \frac{\lout^2}{r^2} \right].
\end{equation}
In the wrinkled region $\lin < r < \lout$, the hoop stress collapses, $\sigma_{rr} \sim 1/r$, and matching yields
\begin{equation}
\sigma_{rr} = \frac{\rf}{\lstar} \frac{2 \lout}{r} \frac{1}{\left(\rf/\lstar\right)^2 + \lout^2}, \quad  \sigma_{\theta\theta} \approx 0.
\end{equation}
In $r < \lin$, we estimate the stress $\sigma_{rr} \sim (\partial w/\partial r)^2$ using $\lin$ as the natural lengthscale, and match with the stress in the wrinkled region at $r = \lin$ to obtain a relationship between $\lin$, $\lout$ and $t$.  
We use the boundary conditions~\eqref{eq:kinematic} and \eqref{eq:vertforce} to eliminate $\lin^3 \sim \sigma_{rr}(\lin) t^2$ and determine that the time at which wrinkles reach the sheet edge (i.e.\,$\lout \sim \rf/\lstar$) is given by
\begin{equation} \label{eq:tedge}
t_{\text{edge}} \sim \left( \frac{\sigmastar}{E\h} \right)^{2/3};
\end{equation}
for the parameters in Table~\ref{table:params} this is $t_{\text{edge}} = O(10^{-2})$, corresponding to a dimensional time on the order of $100\,\mathrm{\mu}\mathrm{s}$ and indentation depth on the order of $100\,\mathrm{\mu}\mathrm{m}$.  This explains why experiments~\cite{pnas} report the sheet being covered in wrinkles from very soon after the beginning of impact.  
We also note that the corresponding dimensional indentation depth $\depth \sim (\sigmastar/E\h)^{2/3} \rf^{2/3} \lstar^{1/3}$, which is different to the static case (cf.~\S\ref{sec:static}).

Finally, we ask how big the tensile zone $r < \lin$ is and whether it affects the bulk solution.  In $r > \lin$ the stress is simply $\sigma_{rr} = 1/r$, $\sigma_{\theta \theta} \approx 0$, and we expect elastocapillary waves to propagate according to $r_m \sim t^{1/2}$.  
Imposing continuity of radial stress at $r = \lin$ then suggests
\begin{equation} \label{eq:li}
\lin \sim \frac{\sigmastar}{E \h} t^{-1},
\end{equation}
so the tensile core is small and shrinking at the times of interest, for example at $t = t_{\text{edge}} \approx 10^{-2}$ we already have $\lin \approx 10^{-1}$.  

Inside the tensile core, we briefly assess how the velocity potential and surface height behave, and whether they change significantly.  Starting with the velocity potential, we note that $\varphi$ and $\partial \varphi/\partial r$ should be continuous across $r = \lin$, and $\partial \varphi/\partial r = 0$ at $r = 0$.  Since the velocity potential is governed by Laplace's equation with a vertical lengthscale set by the bulk motion, we infer that $\varphi$ and $\partial \varphi/\partial r$ do not change across the short tensile zone, and that it is appropriate to apply a boundary condition $\partial \varphi/\partial r = 0$ to the bulk solution as $r \rightarrow 0$.  At the interface, the dynamic boundary condition may be rewritten $r \sigma_{rr} (\partial w/\partial r) = \text{constant} = O(1)$, so $w$ does not change an $O(1)$ amount over a distance $\lin \ll 1$.  
We therefore expect the problem in the wrinkled bulk to be largely unaffected by the tensile core.

\section{Role of finite impactor radius} \label{sec:finitealpha}

In \S\ref{sec:bareimpact}, we described previous work on impact at a bare liquid--air interface, resulting in self-similar behaviour with radial extent $r \sim t^{2/3}$ for $\weber \ll 1$ and $r \sim t^{1/2}$ for $\weber \gg 1$.  In \S\ref{sec:intro}, we outlined a mathematical model for the impact of a point impactor onto a thin elastic sheet floating at a liquid--air interface (presenting a self-similar solution with lateral scale $r \sim t^{1/2}$ in \S\ref{sec:elastocapillarywave}).  In this section, we investigate the role of a finite impactor radius on the resulting dynamics and see that impactor radius enters the problem as a modified Weber number.  We shall see below that, for impact on a floating sheet, the lateral scale $r \sim t^{1/2}$ for any modified Weber number, allowing us to smoothly connect the limiting cases in which surface tension dominates or is negligible. 

\subsection{Mathematical model} \label{sec:finitealpha:model}
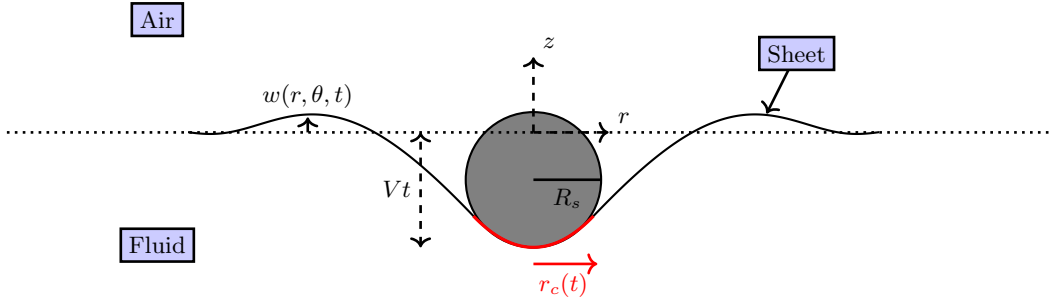
\begin{figure}[ht]
\begin{tikzpicture}
\fill[circle,gray] (0,-0.63) circle(0.9cm);
\draw[circle,thick] (0,-0.63) circle (0.9cm);
\draw (0,-0.63) to (0.9cm,-0.63);
\node [below] at (0.45cm,-0.63) {$\rs$};

\draw[{<[scale=1.0]}-{>[scale=1.0]},dashed] (-1.5,-0.68-0.85) to (-1.5,0.0);
\node[left] at (-1.5,-0.75) {$Vt$};

\node[black] at (-5.0,-1.5)[rectangle,draw,fill=blue!20] {{Fluid}};
\node[black] at (-5.0,1.5)[rectangle,draw,fill=blue!20] {{Air}};
\draw[<-] (3.1,0.25) to (3.5,1.05) node[rectangle,draw,fill=blue!20] {Sheet};

\draw[->,dashed] (0,0) to (0.0,1.0) node[above right]{$z$};
\draw[->,dashed] (0.0,0.0) to (1.0,0.0) node [above right] {$r$};
\draw[dotted] (-7,0.0) to (7,0.0);

\draw[->] (-3.0,0.0) to (-3.0,0.2);
\node[above] at (-3.0,0.2) {$w(r,\theta,t)$};

\draw[->,red] (0.0,-1.75) to (0.85cm,-1.75);
\node[below,red] at (0.4cm,-1.75) {$\rc(t)$};

\begin{axis}[
    at = {(0,0)}, anchor = {origin},
    width = 0.6\textwidth,
    height = 0.2\textwidth,
    xmin = -3, xmax = 3,
    ymin = -1.01, ymax = 0.3,
    every axis plot/.append style={thick},
    axis line style={draw=none},
    ticks=none,
    ]
    \addplot[black] table {similarity-solution/intro/profile_epsilon2.txt};
    \addplot[red,very thick] table {similarity-solution/intro/contact_epsilon2.txt};
\end{axis}

\end{tikzpicture}
{\caption{Schematic diagram: a sphere of radius $\rs$, with velocity $-\V\unitk$, impacts a floating elastic sheet, causing a vertical deflection of the sheet $w(r,\theta,t)$.  The contact region $r\leq r_c(t)$ between the sphere and sheet is highlighted in red. }
\label{fig:model}}
\end{figure}

At time $t=0$, a  sphere of radius $\rs$  and   velocity $\boldsymbol{v}=-\V\unitk$ impacts the sheet, as illustrated in Figure~\ref{fig:model}. The region in which the sheet is in contact with the sphere evolves dynamically and so is written $0\leq r\leq \rc(t)$. We approximate the sphere's surface by a paraboloid --- an approximation that is valid provided that $\rc \ll \rs$, and so is expected to hold at early times. In the contact region (in dimensional terms) we immediately have
\begin{equation} \label{eq:paraboloid}
w(r,t) = - V t + \frac{r^2}{2 \rs}, \quad 0\leq r<\rc(t).
\end{equation} The contact radius $\rc(t)$ is not known \emph{a priori} and so must be determined as part of the solution of the problem. 
Requiring the surface height and the vertical fluid velocity to be continuous across the contact line is equivalent to assuming that the sheet height, $w$, and slope, $\partial w/\partial r$, are both continuous at $r = \rc(t)$.   
We shall also assume that the sheet wraps the sphere without  stretching significantly, so that $u_r = -r^3/(6 \rs^2)$ and $u_{\theta} = 0$ {in $r\leq\rc(t)$}.  We note that, alternatively, the sheet may wrap the sphere without pulling additional material radially inward, so that $u_r = u_{\theta} = 0$ in $r \leq \rc(t)$. We find below that these different boundary conditions are experimentally indistinguishable.

We non-dimensionalize and derive the same dimensionless governing equations as before; now, however the point impact conditions at $r = 0$ are replaced by new equations describing the deformation inside the contact region $r \leq \rc$.  In line with the asymptotic expansions in \S\ref{sec:highbendabilitylimit}, we expand $\rc$ in powers of $\delta$, with $\rca$ denoting the leading-order contact position in the limit $\spar \rightarrow 0$.  At leading order in the high-bendability limit $\spar \rightarrow 0$, the conditions~\eqref{eq:defcontact0} are replaced by
\begin{equation} \label{eq:def2}
\bw{0} = -t + \frac{r^2}{2\simpar}, \quad
\bur{0} = - \frac{r^3}{6\alpha^2}, \quad \but{0} = 0,
\end{equation}
throughout $0<r\leq\rca(t)$ where
\begin{equation}
\simpar = \frac{\rs}{\lstar} = \frac{\rho V^2 \rs}{\sigmastar}
\label{eqn:alphaDefn}
\end{equation} 
is the dimensionless sphere radius and can be interpreted as a modified Weber number, with the impactor radius $\rs$ providing the appropriate lengthscale and $\sigmastar = \gamma \rf/\lstar$ the typical tension.  We note that, for the experiments presented in~\cite{pnas}, $\simpar$ lies in the range $ 0.34\lesssim \simpar\lesssim4.65$; we shall see that this finite-$\simpar$ model is required to accurately describe the observed behaviour. 

We require the surface height and its slope to be continuous at the edge of the contact region, so
\begin{equation} \label{eq:contact}
\bw{0} = -t + \frac{(\rca)^2}{2\simpar}, \quad \frac{\partial \bw{0}}{\partial r} = \frac{\rca}{\simpar} \quad \text{at} \quad r = \rca.
\end{equation}
We now apply the dynamic boundary condition~\eqref{eq:simdyn} in $r > \rca$ only.

We propose a similarity solution of the same form as before, but with an additional variable
\begin{equation}
\rca = t^{1/2} \con_0. 
\end{equation}
In similarity variables, the kinematic boundary condition~\eqref{eq:kinematic0} is replaced with
\begin{equation} \label{eq:finitalphakinematic}
\frac{\partial \f}{\partial \zeta} = \left\{ \begin{array}{cc}
- 1, & \xi \leq \con_0
\\
\Hsim - \tfrac{1}{2} \xi \tfrac{\mathrm{d} \Hsim}{\mathrm{d} \xi}, & \xi > \con_0
\end{array} \right.,
\end{equation}
and the dynamic boundary condition is applied in $\xi > \con_0$. 
Imposing the condition of continuity of the surface and its radial derivative at $\xi = \con_0$, \eqref{eq:contact}, we find
\begin{equation} \label{eq:continuity}
\Hsim(\con_0) = - 1 + \frac{(\con_0)^2}{2\simpar}, \quad \left.\frac{\mathrm{d} \Hsim}{\mathrm{d} \xi}\right|_{\con_0} = \frac{\con_0}{\simpar}.
\end{equation}

\begin{figure}[ht]
\begin{tikzpicture}
\begin{axis}[
    at = {(-0.02\textwidth,0.05\textwidth)}, anchor = {south east},
    width = 0.33\textwidth,
    height = 0.22\textwidth,
	xmin=0.01, xmax=10,
	ymin=0.01, ymax=10,
	x label style={at={(current axis.below origin)},anchor=center, below=5mm,right=0.11\textwidth},
    y label style={at={(current axis.above origin)},rotate=-90,anchor=center,left=9mm,below=0.045\textwidth},
	xlabel = {$\simpar$},
	ylabel = {$\con_0$},
	every axis plot/.append style={very thick},
	xmode=log,ymode=log,
	minor tick style={draw=none},
]
	\addplot [red] table {similarity-solution/similarity_alpha_delta_-2to1.txt};
	\addplot [black, dashed] table {similarity-solution/wagner.txt};
	\addplot [black, thick] table {similarity-solution/wagner_vert.txt};
	\addplot [black, thick] table {similarity-solution/wagner_horz.txt};
	\node[scale=0.75] at (0.2,1.1) {1};
	\node[scale=0.75] at (1.2,1.9) {2};
	
	\addplot[black, dashed] table {similarity-solution/pointindenter.txt};
	\addplot[black] table {similarity-solution/pointindenter_horz.txt};
	\addplot[black] table {similarity-solution/pointindenter_vert.txt};
	\node[scale=0.75] at (-3.8,-2.8) {1};
	\node[scale=0.75] at (-2.7,-1.6) {1};
\end{axis}
\begin{axis}[
    at = {(0,0)}, anchor = {south east},
    width = 0.7\textwidth,
    height = 0.4\textwidth,
    xmin = 0, xmax = 8,
    ymin = -1, ymax = 0.5,
    xlabel = {$\xi$},
    ylabel = {$H(\xi)$},
    every axis plot/.append style={very thick},
    ]
    \addplot[blue,dashed] table {similarity-solution/profiles-prf/contactonly-epsilon0.5.txt};
    \addplot[blue] table {similarity-solution/profiles-prf/outeronly-epsilon0.5.txt};
    \addplot[blue,only marks,mark=*] table {similarity-solution/profiles-prf/firstmin-epsilon0.5.txt};
    \addplot[violet,dashed] table {similarity-solution/profiles-prf/contactonly-epsilon1.txt};
    \addplot[violet] table {similarity-solution/profiles-prf/outeronly-epsilon1.txt};
    \addplot[violet,only marks,mark=*] table {similarity-solution/profiles-prf/firstmin-epsilon1.txt};
    \addplot[red,dashed] table {similarity-solution/profiles-prf/contactonly-epsilon2.txt};
    \addplot[red] table {similarity-solution/profiles-prf/outeronly-epsilon2.txt};
    \addplot[red,only marks,mark=*] table {similarity-solution/profiles-prf/firstmin-epsilon2.txt};
    \addplot[black] table {similarity-solution/profiles/pointindenterprofile.txt};
    \addplot[black, only marks, mark=*] table {similarity-solution/profiles/pointindenterminimum.txt};
    
    \draw[{<[scale=1.3]}-,black] (0.065\textwidth,0.145\textwidth)--(0.04\textwidth,0.20\textwidth) node[above, text width=20mm,align=center] {point impactor};
    \draw[-{>[scale=1.3]},black] (0.055\textwidth,0.11\textwidth)--(0.12\textwidth,0.07\textwidth) node[below right]{increasing $\simpar$};
    \draw[{<[scale=1.3]}-,black] (0.215\textwidth,0.19\textwidth)--(0.215\textwidth,0.15\textwidth) node[below, text width=30mm,align=center] {experimental wavefront};
\end{axis}
\end{tikzpicture}
{\caption{Numerical solution of the early-time similarity problem~\eqref{eq:laplace4}--\eqref{eq:farfield} showing the similarity profile of the deformed surface for a point impactor (black), and of the early-time finite-radius problem~\eqref{eq:laplace4}, \eqref{eq:simdyn}, \eqref{eq:farfield}, \eqref{eq:finitalphakinematic} and~\eqref{eq:continuity}  for dimensionless sphere radii $\simpar = 0.5$ (red), $1$ (purple), $2$ (blue). For the finite-radius impactor ($\simpar>0$), dashed curves show the region in which the sheet is in contact with the impacting sphere; the location of the wavefront (defined as the first local minimum of the interface beyond the contact in~\cite{pnas}) is indicated by a large circle.  Inset: the similarity coordinate of the contact point, $\con_0 = r_c/t^{1/2}$, as a function of dimensionless sphere radius $\simpar$.  The dashed lines show the point impactor prediction $\xi_c {\approx} 0.86\simpar$, valid for $\simpar\ll1$, and the $\alpha \rightarrow \infty$ prediction $\xi_c = \sqrt{3 \simpar}$, valid for $\simpar\gg1$~\cite{Philippi2016}.}
\label{fig:similarity}}
\end{figure}
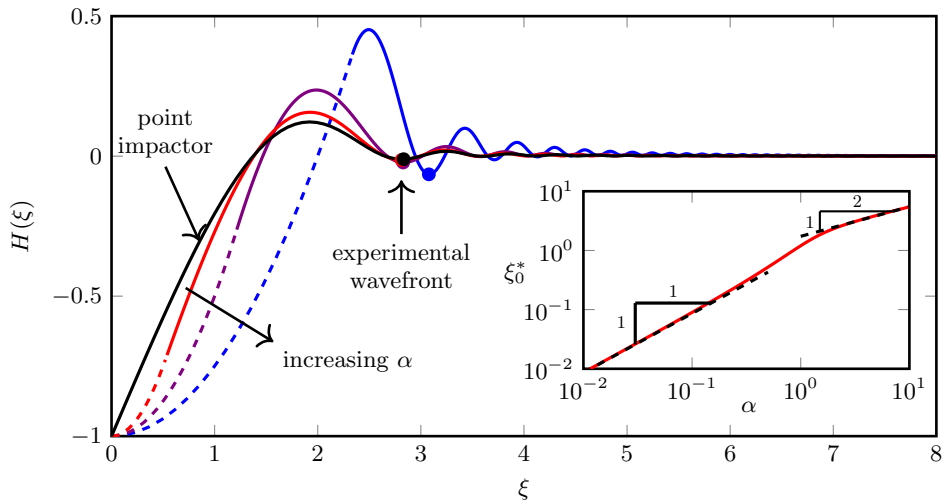

The new similarity problem was again solved numerically using second-order finite differences for different values of the dimensionless sphere radius, $\simpar = \rs/\lstar$  \eqref{eqn:alphaDefn}, which  is the only remaining parameter in  this leading-order problem. Our numerical solution uses a non-uniform grid spacing to provide fine resolution near the impactor, with a coarser resolution used in the far field.  An initial guess was provided for the contact point $\con_0$, which is not known \emph{a priori} and must be determined as part of the solution. A nonlinear solver (\texttt{fsolve} in Matlab) was used to determine the value of $\con_0$ for which the slope of the surface satisfied the smoothness condition \eqref{eq:continuity}.  The free surface is illustrated in Figure~\ref{fig:similarity} for different values of $\simpar$.
 
Our numerical results show that the position of the contact point in similarity variables, $\con_0$, increases as the dimensionless radius $\simpar$ increases (see inset of Figure~\ref{fig:similarity}).  The limit $\simpar\to\infty$ corresponds to zero interfacial tension (and $\weber \rightarrow \infty$) and so should be expected to reduce to the classic impact problem in the absence of interfacial tension \cite{Wagner1932,Korobkin1988,Howison1991,Philippi2016}; in particular, recasting results from the classic Wagner problem in our notation we expect that  $\con_0 \sim \sqrt{3 \simpar}$ as $\simpar \rightarrow \infty$. This is consistent with our  numerical results.  Conversely, in the limit $\alpha \rightarrow 0$ we recover the point indenter case outlined in \S\ref{sec:elastocapillarywave}.  The size of the contact region can be approximated for $\alpha \ll 1$ by matching the predicted slope of the free surface at $\xi = 0$ with an inner region $\xi$, $\zeta = O(\simpar)$ in the neighbourhood of the impact point (where surface tension dominates the interfacial stress balance~\eqref{eq:simdyn} and the interface deflection is linear). This rescaling shows that   $\con_0/\simpar\approx0.86$ for $\simpar\ll1$, again consistent with our numerics.

\subsection{Comparison with experiments}
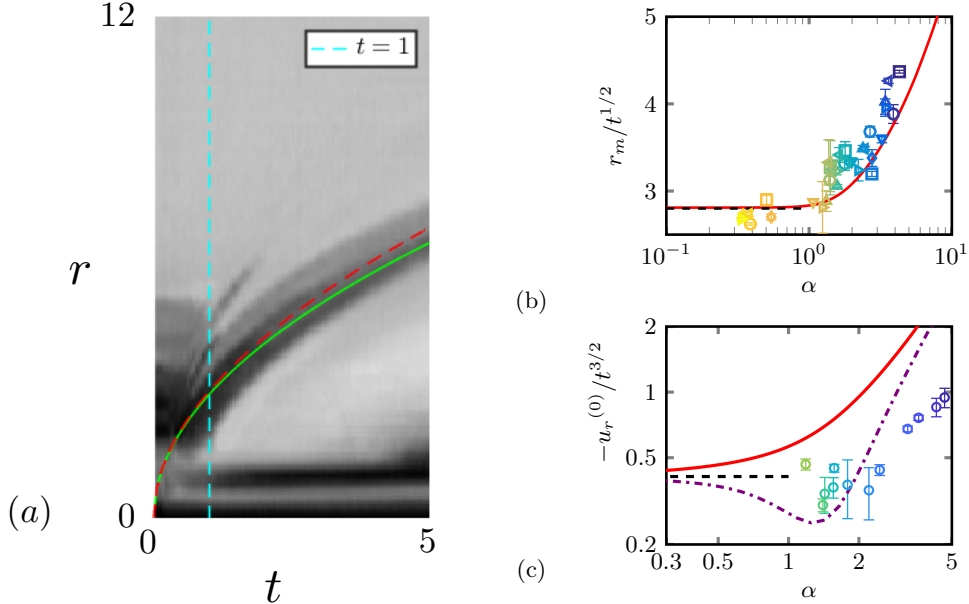
\begin{figure}[b]
\begin{tikzpicture}
\begin{axis}[
    at = {(0.0\textwidth, 0.08\textwidth)},
    width = 0.3\textwidth,
    height = 0.25\textwidth,
    xmode = log,
    xlabel = {$\simpar$},
    ylabel = {$r_m/t^{1/2}$},
    xmin = 0.1, xmax = 10,
    ymin = 2.5, ymax = 5.0,
    major tick style = {very thick},
    major tick length = {1mm},
    cycle list name = mylist,
    ]
    \foreach \k in {1, ..., 33} 
        {
            \addplot+[only marks, error bars/.cd, y dir = both, y explicit, error bar style = {line width = 0.5pt},error mark options={ rotate=90, line width=0.5pt}]     table[x index = 0, y index = 1, y error index = 2] {prldata/brokenupdata3/finndata\k.txt};
        }
        
    \addplot[red] table {prldata/firstminforpaper.txt};
    \addplot[black,dashed] table {prldata/asymptotics.txt};
\end{axis}
\node at (-0.32\textwidth,+0.03\textwidth){\includegraphics[width=6.0cm,trim=0 0 12.5cm 0,clip]{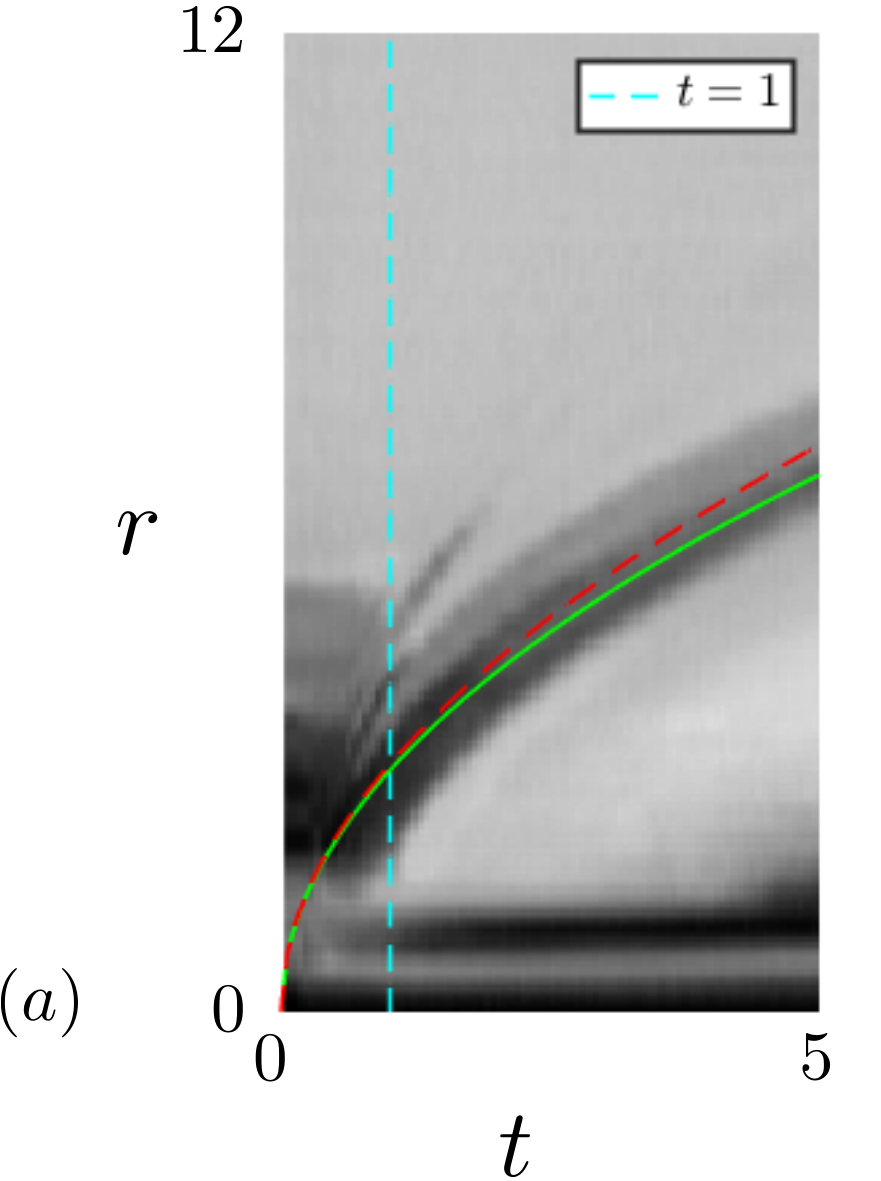}};
\begin{axis}[
    at = {(0,-0.15\textwidth)}, 
    width = 0.3\textwidth,
    height = 0.25\textwidth,
    xmode=log, ymode=log,
    xmin = 0.3, xmax = 5,
    ymin = 0.2, ymax = 2,
    xlabel = {$\simpar$},
    ylabel = {$-\ur{0}/t^{3/2}$},
    every axis plot/.append style={very thick},
    major tick style = {very thick},
    major tick length = {1mm},
    y label style={at={(current axis.above origin)},anchor=center,left=10mm,below=-0.065\textwidth},
    cycle list name = list2,
    xtick={0.3,0.5,1,2,5},
    xticklabels = {0.3,0.5,1,2,5},
    ytick = {0.2, 0.5, 1, 2},
    yticklabels = {0.2, 0.5, 1, 2},
    ]
    \foreach \k in {1, ..., 12}
    {
    \addplot+[only marks, mark = o, mark options = {fill = none, scale = 0.8, line width = 0.8},
                            error bars/.cd, y dir = both, y explicit, error bar style = {line width = 0.5pt}, error mark options = {rotate = 90, line width = 0.5pt}]  table[x index = 0, y index = 1, y error index = 2] {prldata_retraction/brokenupdata4/data\k.txt};
    }
    
    \addplot[red] table {similarity-solution/retraction/edgeprediction_full_newX2_from0.txt};
    \addplot[violet, dash dot] table {similarity-solution/retraction/edgeprediction_full_newX2_fromcontact.txt};
    \addplot[black,dashed] table {similarity-solution/retraction/pointindenter_edgeretraction.txt};
\end{axis}
\node at (-0.1\textwidth,+0.03\textwidth){(b)};
\node at (-0.1\textwidth,-0.17\textwidth){(c)};
\end{tikzpicture}
{\caption{Comparison between theoretical predictions and experimental observations of the propagation of the transverse wave formed when a floating elastic sheet is impacted by a solid sphere: (a) spatio-temporal diagram from an experiment with $R_f=17.35$\,mm, $\h=350$\,nm, $R_s =1.25$\,mm and $V = 1.57$\,ms$^{-1}$, described in~\cite{pnas}.  A radial slice  is imaged over a $2\mathrm{~ms}$ time interval (at 39,024 frames per second), resulting in a spatio-temporal plot.  The first minimum $r_m$ in the propagating transverse wave can be seen as a dark curve, the leading-order early-time prediction from the numerical solution of the similarity problem~\eqref{eq:laplace4}--\eqref{eq:farfield} is shown by the solid green curve and the two-term solution is shown by the red dashed curve. (b) The experimentally measured prefactor in the wavefront propagation $r_m \sim t^{1/2}$ is plotted as a function of dimensionless sphere radius $\simpar$.  The solid red curve shows the prediction from the similarity solution described in \S\ref{sec:simsol}, while the data points are extracted from the data in Figure 3(a) of ref.~\cite{pnas}.  (Colours and symbols are chosen to match those used in~\cite{pnas}.)  The black dashed line shows the point impactor prediction $r_m/t^{1/2} \approx 2.8$.  (c) The prefactor in the edge retraction $-u_r \sim t^{3/2}$ as a function of dimensionless sphere radius $\simpar$.  The similarity solution computed from~\eqref{eq:retraction2} is shown by the solid red curve, and the prediction for $u_r(\con) = 0$ is shown by the violet dash-dotted curve. 
The point impactor prediction $u_r \approx -0.41t^{3/2}$ is shown by the black dashed line.  The data points are extracted from edge retraction measurements in~\cite{pnas} (see Figure 3b of that paper), and the colours here are chosen to match those used there (indicating the value of $\alpha$). In both (b) and (c) the error bars represent a 95\% confidence interval on the fit to data.}
\label{fig:similarity+expts}}
\end{figure}

We now compare the predictions of our model with data from an experimental study (described in~\cite{pnas}) in which a sphere impacts a floating elastic sheet.  
In Figure~\ref{fig:similarity+expts}(a) we show a spatio-temporal plot of the transverse wave compiled during impact: consecutive images from a particular radial line in the sheet are printed side by side, to give time increasing from left to right.  Experimentally, the most  easily identifiable feature in the transverse wave is a dark region corresponding to the first minimum of the interface profile beyond the contact region.  The wavefront location predicted by the similarity solution outlined above is overlaid on this spatio-temporal plot, and gives a good account of the experimentally observed position of the front, even beyond early times.  

A more detailed comparison between the predicted wavefront position and the experimental measurements of~\cite{pnas} is performed by fitting experimental data to a curve $r_m \sim t^{1/2}$ and extracting the prefactor for different experimental parameters (and hence different dimensionless sphere radii $\alpha$).  
The results of this procedure are shown in  Figure~\ref{fig:similarity+expts}(b) and show that the trend predicted by the model agrees well with the experiments.  Although $\simpar = O(1)$ in the experiments of~\cite{pnas}, we note that the point impactor analysis $\simpar \rightarrow 0$ (discussed above) suggests that the predicted position of the wavefront in similarity variables tends to a fixed position $\xi = 2.8$ as $\alpha \rightarrow 0$; moreover this is a good approximation for $\simpar < 1$.

\pgfplotscreateplotcyclelist{mylist}{%
{col1,mark=square, mark options = {fill=none,scale=1.0} },
{col2,mark=o, mark options = {fill=none,scale=1.0} },
{col3,mark=triangle, mark options = {fill=none,scale=1.0, rotate = 90}},
{col4,mark=triangle, mark options = {fill=none,scale=1.0} },
{col5,mark=triangle,mark options={fill=none,scale=1.0, rotate=-90}},
{col6,mark=triangle,mark options={fill=none,scale=1.0, rotate=180}},
{col7,mark=diamond, mark options = {fill=none,scale=1.0} },
{col8,mark=square, mark options = {fill=none,scale=1.0} },
{col9,mark=o, mark options = {fill=none,scale=1.0} },
{col10,mark=triangle,mark options={fill=none,scale=1.0, rotate=90}},
{col11,mark=triangle, mark options = {fill=none,scale=1.0} },
{col12,mark=triangle,mark options={fill=none,scale=1.0, rotate=-90}},
{col13,mark=triangle,mark options={fill=none,scale=1.0, rotate=180}},
{col14,mark=diamond, mark options = {fill=none,scale=1.0} },
{col15,mark=square, mark options = {fill=none,scale=1.0} },
{col16,mark=o, mark options = {fill=none,scale=1.0} },
{col17,mark=triangle,mark options={fill=none,scale=1.0, rotate=90}},
{col18,mark=triangle, mark options = {fill=none,scale=1.0} },
{col19,mark=triangle,mark options={fill=none,scale=1.0, rotate=-90}},
{col20,mark=triangle,mark options={fill=none,scale=1.0, rotate=180}},
{col21,mark=diamond, mark options = {fill=none,scale=1.0} },
{col22,mark=square, mark options = {fill=none,scale=1.0} },
{col23,mark=o, mark options = {fill=none,scale=1.0} },
{col24,mark=triangle,mark options={fill=none,scale=1.0, rotate=90}},
{col25,mark=triangle, mark options = {fill=none,scale=1.0} },
{col26,mark=triangle,mark options={fill=none,scale=1.0, rotate=-90}},
{col27,mark=triangle,mark options={fill=none,scale=1.0, rotate=180}},
{col28,mark=diamond, mark options = {fill=none,scale=1.0} },
{col29,mark=square, mark options = {fill=none,scale=1.0} },
{col30,mark=o, mark options = {fill=none,scale=1.0} },
{col31,mark=triangle,mark options={fill=none,scale=1.0, rotate=90}},
{col32,mark=triangle, mark options = {fill=none,scale=1.0} },
{col33,mark=triangle,mark options={fill=none,scale=1.0, rotate=-90}},
{col34,mark=triangle,mark options={fill=none,scale=1.0, rotate=180}},
}

\pgfplotscreateplotcyclelist{list2}{
{val1},{val2},{val3},{val4},{val5},{val6},{val7},{val8},{val9},{val10},{val11},{val12},{val13},{val13},{val14},{val15}
}

In both of the comparisons presented in Figure~\ref{fig:similarity+expts}(a) and (b) we observe that the model slightly under-predicts the speed of the propagating wave for $t < 1$.  One possible source of error is the increasing importance of the nonlinear terms in~\eqref{eq:kinematic-lo-a} and \eqref{eq:dynlo-a} as time increases.  This motivates determining the higher-order corrections to the similarity solution, i.e.\ the terms $H_1(\xi)$ etc.\ in~\eqref{eq:selfsim}.  We give details of this calculation in Appendix~\ref{sec:simsolcorr}: ultimately, a correction to the location of the wavefront may be determined for each value of the dimensionless parameter $\simpar$.  
A two-term prediction for the location of the first minimum (i.e.\  the wavefront) is shown in Figure~\ref{fig:similarity+expts}(a).  
This correction to the leading-order similarity solution suggests that the nonlinear terms that become non-negligible at later times cause the wave to spread out slightly more rapidly.

We can calculate the edge retraction using the integral
\begin{equation} \label{eq:retraction2}
\bur{0} = 
\left[ 
- \left( \frac{(\con)^3}{6 \simpar^2} \right) 
- \frac{1}{2} \int\limits_{\con}^{\xi} \left( \frac{\mathrm{d} \Hsim}{\mathrm{d} \xi'} \right)^2\ \mathrm{d} \xi' 
\right] 
t^{3/2},
\end{equation}
for the perfect wrapping case, with the additive factor $(-\xi^{*3}/6\simpar^2)$ replaced by zero for the alternative, perfect no-slip model presented in \S\ref{sec:finitealpha:model}.  
We observe that the prefactor in the $t^{3/2}$ power law depends on the dimensionless sphere radius $\simpar$, as illustrated in Figure~\ref{fig:similarity+expts}(c).  
We observe that in both models the prefactor increases with increasing $\simpar > 1$, and, as should be expected, in the $\simpar \rightarrow 0$ limit recovers point-impactor behaviour.  
The predicted edge retraction is compared to the experimental measurements of~\cite{pnas} by fitting experimental data for $t < (\simpar/\con)^2$ to a curve $u_r \sim -t^{3/2}$ and extracting the prefactor for different experimental parameters (see Figure~\ref{fig:similarity+expts}c).  Although the order-of-magnitude predictions for the prefactor are reasonable, the experimental measurement of the retraction of the edge is difficult, and the level of accuracy offered by these experiments does not allow us to distinguish the two contact models (perfect wrapping or perfect no-slip).

\subsection{Validity of similarity solution}
Finally, we check when the similarity solution is valid.  
As noted above, we have assumed that $t \ll 1$ in the above analysis, allowing some terms to be neglected. Further, our parabolic approximation of the impactor requires that $\rc \ll \rs$, which in dimensionless terms becomes $t \ll (\simpar /\con_0)^2$. The determined similarity solution is valid when both conditions are satisfied. For the experiments reported in \cite{pnas},  $0.65\lesssim(\simpar/\con_0)^2\lesssim1.5$.  

We also note that experimentally the velocity of the impactor is not controlled throughout, but rather a sphere impacts the sheet at a known speed and may subsequently decelerate due to resisting forces of the sheet and water.  We account for these effects in Appendix~\ref{sec:deceleration} and find that they do not affect the leading-order result for $t \ll 1$, nor do they affect the first correction mentioned above.  Nonetheless, deceleration may become significant at later times for small spheres with less inertia; eqn~\eqref{eq:app:decelerationtime} provides an upper bound on the times at which our constant-velocity approximation is valid.

\section{Wrinkles} \label{sec:wrinkles}

We now turn our attention to the formation and evolution of wrinkles in the floating elastic sheet during impact.

\subsection{$ t = O(1)$ timescale}
We begin our study of wrinkle formation by proceeding to next order in the governing equations~\eqref{eq:laplace2}--\eqref{eq:farfieldall}.  In particular, evaluating the oscillatory component of the vertical force balance~\eqref{eq:vertforce} at $O(\spar)$, we find that
\begin{equation} \label{eq:wquasi}
\frac{1}{r^4} \frac{\partial^4 \R{w}^{(1)}}{\partial \btheta^4} - \frac{1}{r} \frac{\partial \R{w}^{(1)}}{\partial r^2} + \left( \frac{E\h}{\sigma^{\ast}_{rr}} \right) \left( \frac{\partial^2 \bar{w}^{(0)}}{\partial r^2} \right)^2 \R{w}^{(1)} - \stt{2} \frac{1}{r^2} \frac{\partial^2 \R{w}^{(1)}}{\partial \btheta^2} = 0,
\end{equation}
where we have used
\begin{equation}
\R{\sigma}_{rr}^{(1)} = - \left( \frac{E\h}{\sigma_{\ast}^{rr}} \right) \frac{\partial^2 \bar{w}^{(0)}}{\partial r^2} \hat{w}^{(1)},
\end{equation}
which is calculated using the stress--strain equations~\eqref{eq:stressstrainrr} and \eqref{eq:stressstrainrt}.  
The unknown hoop stress is given by ensuring the wrinkles satisfy the length constraint
\begin{equation} \label{eq:lengthquasi}
\overline{ \left[ \frac{1}{2} \left( \frac{1}{r} \frac{\partial \hat{w}^{(1)}}{\partial \btheta} \right)^2 \right] } = - \frac{\bur{0}}{r},
\end{equation}
which has been derived from~\eqref{eq:stressstraintt}, and the leading-order radial retraction $\bur{0}$ is given by~\eqref{eq:retraction}. 

The equations~\eqref{eq:wquasi} and~\eqref{eq:lengthquasi} simply describe the quasi-static wrinkle profile of a thin sheet under an applied radial retraction --- see e.g.~\cite{Paulsen2016pnas}.  We recognise the first term in~\eqref{eq:wquasi} as a bending term, which opposes the formation of very fine-scale wrinkles, and the second two terms as corresponding to tension- and curvature-induced stiffnesses respectively, which oppose large vertical deformations and hence balance the bending term.  In the last term, the residual hoop stress $\stt{2}$ simply acts as a Lagrange multiplier for the length constraint~\eqref{eq:lengthquasi}.  We conclude that, on this timescale, we observe wrinkles that evolve as the stiffnesses and excess length evolve, but are in quasistatic equilibrium and unaffected by the inertia of the underlying fluid.  To take account of the effect of fluid inertia, we must consider the behaviour on a shorter timescale.

\subsection{Modelling inertia-modulated wrinkling}
Based on the dominant balance argument presented in~\S\ref{sec:nondim-action}, we study wrinkle formation on the timescale $t = \spar^{1/2} \bt \ll 1$, with $\bt = O(1)$.  Based on our analysis at $t = O(1)$, we anticipate that the leading-order axisymmetric variables will evolve in a self-similar power-law fashion, motivating a rescaling in $\spar$.  
We focus our attention on the portion of the sheet where the hoop stress $\sigma_{\theta\theta}<0$, and the sheet wrinkles.   An in-depth analysis (see Appendix~\ref{app:wrinkles}) shows that the compressive hoop stress is $O(\spar^{3/2})$ near the impact point, and hence larger than the $\sigma_{\theta\theta}=O(\spar^2)$ behaviour observed in static wrinkling problems \cite{Davidovitch2011,Taffetani2017}.  However, in the flat portion of the sheet far from the impact point, $\sigma_{\theta\theta} = O(\spar^2)$, as is more usual in wrinkling problems.  We {focus on this flat region to allow us to} determine a simple system of equations governing these variables in the high-bendability limit $\spar \rightarrow 0$; here we  justify these equations intuitively, while a more formal derivation may be found in Appendix~\ref{app:wrinkles}. 

We derive approximate governing equations for the evolution of the wrinkle profile (and the associated velocity potential) by considering a material circle in the sheet far from the point of impact, so that the sheet is approximately flat, and undergoes a radial retraction $u_r(t)$ given by~\eqref{eq:retraction}.
We confine our attention to radii sufficiently large that the length of the materical circle $2\pi r \gg \spar \lstar$, which is the characteristic wrinkle wavelength scale; we can then `unwrap' this material circle, treating it as a one-dimensional sheet, see Figure~\ref{fig:1Dwrinkles}. 

\definecolor{babyblue}{rgb}{0.54, 0.81, 0.94}

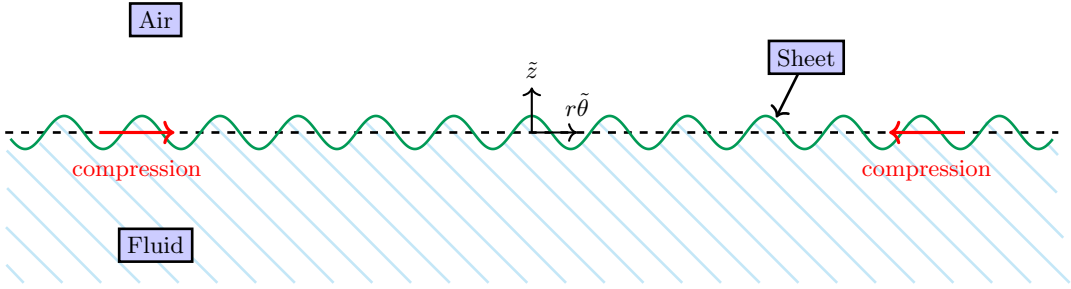
\begin{figure}[ht]
\begin{tikzpicture}

\def\rep{1.03}
\def\sep{0.515}
\def\sepa{0.3}
\def\height{0.14}
\def\depth{-0.14}
\def\bottom{-2.0}
\def\right{2.0}
\def\opac{0.5}

\draw[babyblue,opacity=\opac] (-0.15,\height) to (\right,\bottom);
\draw[babyblue,opacity=\opac] (-0.15+\rep,\height) to (\right+\rep,\bottom);
\draw[babyblue,opacity=\opac] (-0.15+2*\rep,\height) to (\right+2*\rep,\bottom);
\draw[babyblue,opacity=\opac] (-0.15+3*\rep,\height) to (\right+3*\rep,\bottom);
\draw[babyblue,opacity=\opac] (-0.15+4*\rep,\height) to (\right+4*\rep,\bottom);
\draw[babyblue,opacity=\opac] (-0.15+5*\rep,\height) to (\right+5*\rep,\bottom);
\draw[babyblue,opacity=\opac] (-0.15+6*\rep,\height) to (\right+6*\rep-1.2,\bottom+1.2);

\draw[babyblue,opacity=\opac] (-0.15-\rep,\height) to (\right-\rep,\bottom);
\draw[babyblue,opacity=\opac] (-0.15-2*\rep,\height) to (\right-2*\rep,\bottom);
\draw[babyblue,opacity=\opac] (-0.15-3*\rep,\height) to (\right-3*\rep,\bottom);
\draw[babyblue,opacity=\opac] (-0.15-4*\rep,\height) to (\right-4*\rep,\bottom);
\draw[babyblue,opacity=\opac] (-0.15-5*\rep,\height) to (\right-5*\rep,\bottom);
\draw[babyblue,opacity=\opac] (-0.15-6*\rep,\height) to (\right-6*\rep,\bottom);
\draw[babyblue,opacity=\opac] (-0.15-7*\rep+0.38,\height-0.38) to (\right-7*\rep,\bottom);
\draw[babyblue,opacity=\opac] (-0.15-8*\rep+1.4,\height-1.4) to (\right-8*\rep,\bottom);

\draw[babyblue,opacity=\opac] (-0.15+\sepa+\sep,\depth) to (\right+\sep,\bottom);
\draw[babyblue,opacity=\opac] (-0.15+\sepa+\sep+\rep,\depth) to (\right+\sep+\rep,\bottom);
\draw[babyblue,opacity=\opac] (-0.15+\sepa+\sep+2*\rep,\depth) to (\right+\sep+2*\rep,\bottom);
\draw[babyblue,opacity=\opac] (-0.15+\sepa+\sep+3*\rep,\depth) to (\right+\sep+3*\rep,\bottom);
\draw[babyblue,opacity=\opac] (-0.15+\sepa+\sep+4*\rep,\depth) to (\right+\sep+4*\rep,\bottom);
\draw[babyblue,opacity=\opac] (-0.15+\sepa+\sep+5*\rep,\depth) to (\right+\sep+5*\rep-0.6,\bottom+0.6);
\draw[babyblue,opacity=\opac] (-0.15+\sepa+\sep+6*\rep,\depth) to (\right+\sep+6*\rep-0.5-1.2,\bottom+0.5+1.2);

\draw[babyblue,opacity=\opac] (-0.15+\sepa+\sep-\rep,\depth) to (\right+\sep-\rep,\bottom);
\draw[babyblue,opacity=\opac] (-0.15+\sepa+\sep-2*\rep,\depth) to (\right+\sep-2*\rep,\bottom);
\draw[babyblue,opacity=\opac] (-0.15+\sepa+\sep-3*\rep,\depth) to (\right+\sep-3*\rep,\bottom);
\draw[babyblue,opacity=\opac] (-0.15+\sepa+\sep-4*\rep,\depth) to (\right+\sep-4*\rep,\bottom);
\draw[babyblue,opacity=\opac] (-0.15+\sepa+\sep-5*\rep,\depth) to (\right+\sep-5*\rep,\bottom);
\draw[babyblue,opacity=\opac] (-0.15+\sepa+\sep-6*\rep,\depth) to (\right+\sep-6*\rep,\bottom);
\draw[babyblue,opacity=\opac] (-0.15+\sepa+\sep-7*\rep,\depth) to (\right+\sep-7*\rep,\bottom);
\draw[babyblue,opacity=\opac] (-0.15+\sepa+\sep-8*\rep+0.6,\depth-0.6) to (\right+\sep-8*\rep,\bottom);
\draw[babyblue,opacity=\opac] (-0.15+\sepa+\sep-9*\rep+1.62,\depth-1.62) to (\right+\sep-9*\rep,\bottom);

\node[black] at (-5.0,-1.5)[rectangle,draw,fill=blue!20] {{Fluid}};
\node[black] at (-5.0,1.5)[rectangle,draw,fill=blue!20] {{Air}};
\draw[<-] (3.25,0.20) to (3.65,1.00) node[rectangle,draw,fill=blue!20] {Sheet};

\draw[dashed,black,domain=-7:7,samples=10] plot (\x,0);
\draw[ForestGreen,domain=-21:21,samples=800] plot (0.33*\x,{0.22*cos(360*\x/(3.14)});
\draw[->,thick] (0,0) to (0.0,0.6) node[above]{$\bz$};
\draw[->,thick] (0.0,0.0) to (0.6,0.0);
\node[above] at (0.6,0.1) {$r\btheta$};

\draw[red,->,very thick] (-5.75,0) to (-4.75,0);
\node[red] at (-5.25,-0.5) {compression};
\draw[red,->,very thick] (5.75,0) to (4.75,0);
\node[red] at (5.25,-0.5) {compression};

\end{tikzpicture}
{\caption{Schematic of wrinkling in a one-dimensional sheet under confinement.} \label{fig:1Dwrinkles}}
\end{figure}

We denote the wrinkle amplitude by $\W$ and the associated velocity potential by $\Pw$, and complete our unwrapping of a material circle by introducing a new variable $\x = r\btheta$.  For a two-dimensional fluid domain below a one-dimensional sheet, the Laplace equation~\eqref{eq:laplace2} simplifies to
\begin{equation}
\label{eqn:1Dlaplace}
\frac{\partial^2 \Pw}{\partial \x^2} + \frac{\partial^2 \Pw}{\partial \bz^2} = 0,
\end{equation}
 with far-field condition
\begin{equation}
\Pw \rightarrow 0 \quad \text{as} \quad \bz \rightarrow -\infty.
\end{equation}
At early times, we linearize the free surface onto the horizontal plane $\bz = 0$ and simplify convective derivatives to partial derivatives.  The kinematic condition at the free surface is then
\begin{equation}
\frac{\partial \Pw}{\partial \bz} = \frac{\partial \W}{\partial \bt}.
\end{equation}
The dynamic condition must take into account (i) the inertia of the underlying fluid, (ii) the bending stiffness of the sheet, and (iii) the compressive force. The one-dimensional version of the vertical force balance~\eqref{eq:vertforce} is then
\begin{equation}\label{eq:onedimforcebalance}
\frac{\partial \Pw}{\partial \bt} + \frac{\partial^4 W}{\partial \x^4} - \C \frac{\partial^2 \W}{\partial \x^2} = 0,
\end{equation}
where $\C = \stt{2} < 0$ is the stress along the material circle.  We reiterate that, near the centre of the sheet, our analysis suggests that the hoop stress $\sigma_{\theta\theta}$ is larger than the $O(\spar^2)$ result usually obtained for static indentation~\cite{Davidovitch2011,Taffetani2017}; however, far from the point of impact we may neglect axisymmetric motion, so $\sigma_{\theta\theta} \approx \spar^2 \bar{\sigma}_{\theta\theta}^{(2)}$ (see Appendix~\ref{app:wrinkles}).  

The size of the compressive stress $\C$ depends on the confinement imposed, and is determined by coupling the excess length absorbed by the wrinkles to the radial retraction~\eqref{eq:retraction} via the geometric constraint
\begin{equation} \label{eq:constrainta}
\frac{1}{2} \int_{-\pi r}^{\pi r} \left( \frac{\partial W}{\partial \x} \right)^2\ \mathrm{d} \x = - 2 \pi \ur{0}/\spar^{3/4} \sim \bt^{3/2}.
\end{equation} 
We note that $r \gg 1$, and the unwrapped material circle is very long compared to the wrinkle wavelength.  We could alternatively write the system~\eqref{eqn:1Dlaplace}--\eqref{eq:constrainta} in terms of a coordinate $\btheta \in [-\pi, \pi]$, which would suggest that $W \sim \sqrt{r}$ to ensure that the integral in~\eqref{eq:constrainta} is $O(1)$.  In fact, we shall scale out all radial dependence in the solution below.

Since the equations \eqref{eqn:1Dlaplace}--\eqref{eq:onedimforcebalance} are linear, it is natural to Fourier Transform in $x = r\btheta$ so that the solution may be written
\begin{subequations} \label{eq:fourier}
\begin{equation}
\Pw(\x,\bz,\bt) = \int_{-\infty}^{\infty} \frac{1}{|\wno|} \frac{\mathrm{d} \coeff_{\wno}}{\mathrm{d} \bt} \mathrm{e}^{i \wno \x} \mathrm{e}^{|\wno| \bz}\ \mathrm{d} \wno,
\end{equation}
\begin{equation}
\W(\x,\bt) = \int_{-\infty}^{\infty} \coeff_{\wno} \mathrm{e}^{i \wno \x}\ \mathrm{d} \wno.
\end{equation}
\end{subequations}
with $\Pw \rightarrow 0$ as $\bz \rightarrow -\infty$ as required, and the coefficients $a_{\wno}(\bt; r)$ to be determined.   
The vertical force balance then reduces to
\begin{equation} \label{eq:evolution}
\frac{\mathrm{d}^2 \coeff_{\wno}}{\mathrm{d} \bt^2} = - |\wno| \left( \wno^4 + \C \wno^2 \right) \coeff_{\wno}.
\end{equation}

The governing equation \eqref{eq:evolution} must be solved with the unknown variable $\C(\bt)$ determined to ensure that the (transformed) compression constraint  \eqref{eq:constrainta} is satisfied. We note first, however, that the linearity of \eqref{eq:evolution}  means that $\coeff_{\wno}$ can be scaled to absorb any prefactor in the geometric constraint~\eqref{eq:constrainta} without affecting the relative sizes of the coefficients $a_k$. In particular, features such as the dominant wavenumber $\wno_*(t)$, defined as the value of $k$ at which $\coeff_{\wno}$ is maximized, should only be  affected by the form of the power law;  it is therefore sufficient to solve~\eqref{eq:evolution} subject to
\begin{equation} \label{eq:constraint}
\int_{-\infty}^{\infty} \wno^2 |\coeff_{\wno}|^2\ \mathrm{d} \wno = \bt^{\hspace{0.5mm} 3/2},
\end{equation}
and with appropriate initial conditions (an example is given in \S\ref{sec:wrinklexpts} below).  A naive balance of terms in~\eqref{eq:evolution} suggests seeking a similarity solution with dominant wavenumber $\wno_*(t) \sim \bt^{-2/5}$ and $\C \sim \bt^{-4/5}$, but we shall see in \S\ref{sec:wrinklenumber} that this is not correct.  First, though, we validate our model by comparing numerical solutions of~\eqref{eq:evolution}, \eqref{eq:constraint} with experimental observations.

\subsection{Comparison with experiments} \label{sec:wrinklexpts}
\begin{figure}[ht]
\centering
\begin{tikzpicture}
\node at (0,-0.35) {\includegraphics[height=0.315\textwidth,trim=9.9cm 0 0 0,clip]{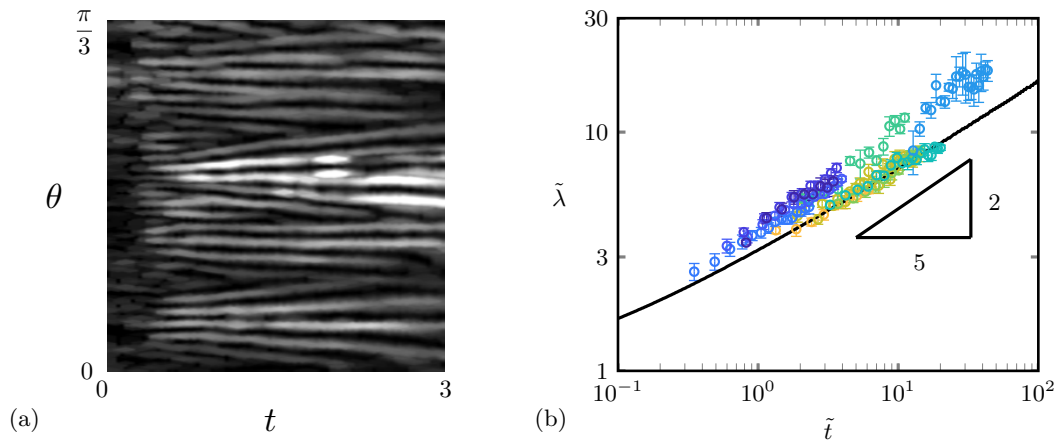}};
\begin{axis}[
    at = {(9.5,0)}, anchor = {center},
    width = 0.4\textwidth,
    height = 0.35\textwidth,
    xmode=log, ymode=log,
    xmin = 0.1, xmax = 100,
    ymin = 1, ymax = 30,
    xlabel = {$\bt$},
    ylabel = {$\tilde{\lambda}$},
    every axis plot/.append style={very thick},
    major tick style = {very thick},
    major tick length = {1mm},
    y label style = {rotate=-90},
    cycle list name = list3,
    xtick={0.1, 1,10,100},
    xticklabels = {$10^{-1}$, $10^{0}$,$10^{1}$, $10^{2}$},
    ytick = {1,  3, 10, 30},
    yticklabels = {1, 3, 10, 30},
    ]
    \foreach \k in {1, ..., 11}
    {
    \addplot+[only marks, mark = o, mark options = {fill = none, scale = 0.8, line width = 0.8},
                            error bars/.cd, y dir = both, y explicit, error bar style = {line width = 0.5pt}, error mark options = {rotate = 90, line width = 0.5pt}]  table[x index = 0, y index = 1, y error index = 2] {wavelength/data\k.txt};
    }
    
    \addplot[black] table {wavelength/numerics.dat};
    \addplot[black] table{wavelength/scaling_slope.dat};
    \addplot[black] table{wavelength/scaling_horz.dat};
    \addplot[black] table{wavelength/scaling_vert.dat};
\end{axis}
    \node at (8.9,-0.9) {5};
    \node at (9.9,-0.1) {2};
\node at (-3,-3) {(a)};
\node at (4,-3) {(b)};
\end{tikzpicture}
{\caption{(a) Spatio-temporal diagram of wrinkles in a floating elastic sheet impacted by a solid sphere.  
An arc of a  material circle (initially at $r=0.8\rf$) is imaged over a $2.5\mathrm{ms}$ interval at 39,024 frames per second: at early times many fine wrinkles are visible (left), but coarsen as time evolves (to the right). These data were obtained for $R_f=13.30$\,mm, $\h=450$\,nm, $R_s =1.25$\,mm and $V = 1.11$\,ms$^{-1}$.  (b) Dominant wavelength $\tilde{\lambda} = 2\pi / \wno_{*}$ of the wrinkle pattern in the flat portion of an impacted sheet, as a function of time.  The data points are a rescaled version of those presented in Figure 4b of~\cite{pnas}, and the colours here are chosen to match.   The solid curve is the numerical solution to \eqref{eq:fourier}--\eqref{eq:constraint} with no fitting parameter, while the triangle illustrates the simple scaling $\tilde{\lambda} \sim \bt^{2/5}$. } \label{fig:wrinklexpts}}
\end{figure}

The experimental study of impact on a floating elastic sheet presented in~\cite{pnas} shows radial wrinkles whose wavelength evolves in time.  An extract from these experiments is shown in Figure~\ref{fig:wrinklexpts}(a), in the form of a spatio-temporal plot in which light and dark regions show the peaks and troughs of the wrinkles, respectively.  At early times, many fine wrinkles are visible, which subsequently merge, so that the wavelength increases (and the number of wrinkles decreases) as time progresses.  This image is taken in the flat part of the sheet, where the static wavelength predicted by~\cite{Paulsen2016pnas} would be governed by a balance between bending stiffness and hydrostatic pressure in the liquid and is given by $\lambda \approx 1.5 \mathrm{mm}$ (see fig.~4c of ref.~\cite{Paulsen2016pnas}).  By constrast, the wavelengths shown in Figure~\ref{fig:wrinklexpts}(a) increase from $\approx 0.3 \mathrm{mm}$ to $\approx 0.6 \mathrm{mm}$.  The important difference, as seen in the vertical force balance~\eqref{eq:onedimforcebalance}, is that here the primary restoring force is the inertia of the underlying fluid, as opposed to hydrostatic pressure.

\pgfplotscreateplotcyclelist{list3}{%
{c11}, {c10}, {c10}, {c9}, {c8}, {c7}, {c5}, {c4}, {c3}, {c2}, {c1}
}

In Figure~\ref{fig:wrinklexpts}(b) we replot experimental data from Figure 4(b) of~\cite{pnas}, showing the dominant wavenumber $\wno_{*}(\bt)$.  We solve our reduced model~\eqref{eq:evolution}--\eqref{eq:constraint} numerically by truncating and discretizing $\wno$ to form a matrix system of the form $M y' = f(y)$.  The mass matrix $M$ is singular because of the inclusion of the integral constraint~\eqref{eq:constraint}, which becomes an algebraic constraint upon discretization.  The resulting matrix system can then be integrated using \texttt{ode23t} in MATLAB, using as initial conditions a uniform initial amplitude and velocity across $k$.  We then determine $\wno_{*}(\bt)$ by identifying the value of $\wno$ for which $|\coeff_{\wno}|$ is maximized at each $\bt$.  Our numerical solution gives reasonable agreement with the experimental measurement, with no fitting parameters (see Figure~\ref{fig:wrinklexpts}b).  
Finally, we note that, since the radial retraction far from the origin is approximately uniform in space, our model predicts that the value of $\wno_*(t)$ should also be independent of radial position, so the typical number of wrinkles in the pattern, $m=r \wno_*(t)/\spar$ should increase with radius.  This feature is also in reasonable agreement with experimental observations (see Figure 4(a) of~\cite{pnas}).

\subsection{Wrinkle evolution} \label{sec:wrinklenumber}
In \S\ref{sec:wrinklexpts} we validated our model by solving the governing equations~\eqref{eq:evolution}--\eqref{eq:constraint} numerically and comparing with experimental observations.  Examining our model in more detail allows us to go beyond the remit of those experiments: we can investigate the distribution of wavelengths in the spectrum~\eqref{eq:fourier}, as well as calculate the residual hoop stress $\sigma_{\theta\theta}^{(2)}$.

First we address the question of whether the equations~\eqref{eq:evolution} and~\eqref{eq:constraint} admit a similarity solution.   For a general retraction power-law (i.e.,~$\sim \bt^{n}$ on the right hand side of~\eqref{eq:constraint}) one might naturally seek a solution of the form
\begin{equation}\label{eqn:SimSoln}
\coeff_{\wno} = \bt^{3/5 + n/2} f(\wnob), \quad  \wno=  \bt^{-2/5} \wnob, \quad \C = \bt^{-4/5} S.
\end{equation} 
We note that the $\wno\sim \bt^{-2/5}$ power law is also found in the interaction between water waves and flexible ice sheets in the ocean, see for example~\cite{Squire1995}.  
However, substituting this ansatz into the evolution equation~\eqref{eq:evolution}, we find that $f \sim \wnob^{-3/2-5n/4}$ as $\wnob \rightarrow 0$; in this case, the integral~\eqref{eq:constraint} does not converge and the geometric constraint cannot be satisfied.  The failure of such a similarity solution is similar to that in the coarsening of wrinkles in a thin sheet on a viscous lubricating film \cite{Kodio2017}.

We proceed to find a numerical solution of the differential-algebraic system~\eqref{eq:evolution} and \eqref{eq:constraint}.  To focus purely on inertia-driven wrinkle coarsening, we neglect the complication of a time-dependent end-shortening by considering the simplest case $n = 0$, so that the right-hand side of~\eqref{eq:constraint} is 1.  For numerical solution over a large range in $\bt$, we must adapt the procedure outlined in \S\ref{sec:wrinklexpts}, because the range of $\wno$ for which $\coeff_{\wno}$ is non-negligible evolves with time: at early times the dominant wave numbers are relatively large, and a correspondingly large range of $\wno$ must be considered. At later times, the dominant wavenumber $\wnostar$ decreases dramatically, necessitating a smaller mesh size to resolve the behaviour at smaller $\wno$ properly. The domain size is decreased correspondingly due to memory contraints --- for example, for $\bt < 1$ the domain is truncated at $\wno = 15$ with mesh size $1/400$ while  for $10^3 < \bt \leq 10^4$ the domain is truncated at $\wno = 2$ with  mesh size  $1/4000$.

\pgfplotsset{
  log y ticks with fixed point/.style={
      yticklabel={
        \pgfkeys{/pgf/fpu=true}
        \pgfmathparse{exp(\tick)}%
        \pgfmathprintnumber[fixed relative, precision=3]{\pgfmathresult}
        \pgfkeys{/pgf/fpu=false}
      }
  }
}

\begin{figure}[ht]
\begin{tikzpicture}
\def\voffset{0.35\textwidth}
\def\offset{1}
\begin{axis}[
    at = {(0,\voffset)}, anchor = {south west},
    width = 0.8\textwidth,
    height = 0.4\textwidth,
    xmin = 0.0, xmax = 3.0,
    ymin = -1, ymax = 1,
    xlabel = {$\wno/\wnostar$},
    ylabel = {$\coeff_{\wno}/\coeff_{\wnostar}$},
    minor tick style={draw=none},
    major tick style = {very thick},
    major tick length = {1mm},
    xtick={0.0, 0.5, 1.0, 1.5, 2.0, 2.5, 3.0},
    ]
    \addplot[orange] table {wavenumber-prf/scaledprofile100.txt};
    \addplot[red] table {wavenumber-prf/scaledprofile10000.txt};
    \addplot[purple] table {wavenumber-prf/scaledprofile1000000.txt};
    \addplot[blue] table{wavenumber-prf/scaledprofile100000000.txt};
    \addplot[ForestGreen] table{wavenumber-prf/scaledprofile10000000000.txt};
\end{axis}
\draw[<-] (5.3,\voffset + 0.12\textwidth) to (5.8,\voffset+0.1\textwidth) node[below right] {increasing $\bt$};
\node at (0.0-\offset,\voffset-\offset) {(a)};

\begin{axis}[
    at = {(0.35\textwidth,0)}, anchor = {south west},
    xmode=log, ymode = log,
    width = 0.45\textwidth,
    height = 0.35\textwidth,
    xmin = 1e-1, xmax =1e10,
    ymin = 1e-4, ymax = 1e1,
    every axis plot/.append style={very thick},
    xlabel = {$\bt$},
    ylabel = $\wnostar$,
    minor y tick style={draw=none},
    major tick style = {very thick},
    major tick length = {1mm},
    ]
    \addplot[blue] table {wavenumber-prf/wavenumber.txt};
    \addplot[black,dashed] table {wavenumber-prf/powerlaw.txt};
\end{axis}

\begin{axis}[
    at = {(0.7\textwidth,0.245\textwidth)}, anchor = {north east},
    width = 0.18\textwidth,
    height = 0.18\textwidth,
    xmin = 0, xmax =1,
    ymin = 2.0, ymax = 2.4,
    ytick={2.0, 2.1, 2.2, 2.3, 2.4},
    xtick={0, 0.5, 1},
    every axis plot/.append style={very thick},
    xlabel = {$\log_{10}[\log_{10}(\bt)]$},
    ylabel = $\wnostar \bt^{2/5}$,
    filter discard warning=false,
    minor tick style={draw=none},
    major tick style = {very thick},
    major tick length = {1mm},
    ]
   \addplot[blue, only marks, mark = *, mark size = 0.15pt] table {wavenumber-prf/compensated2.txt};
\end{axis}
\begin{axis}[
    at = {(0,0)}, anchor = {south west},
    xmode = log, ymode = log,
    width = 0.35\textwidth,
    height = 0.35\textwidth,
    xmin = 1e-1, xmax = 1e8,
    ymin = 1e-6, ymax = 1e1,
    every axis plot/.append style={very thick},
    xlabel = {$\bt$}, ylabel = $-\C$,
    minor tick style={draw=none},
    major tick style = {very thick},
    major tick length = {1mm},
    ]
    \addplot[blue, only marks, mark size = 0.15pt] table {wavenumber-prf/forpaper2.txt};
    \addplot[black, dashed] table {wavenumber-prf/forpaper_powerlaw.txt};
\end{axis}
\node at (0-\offset,0-\offset) {(b)};
\node at (0.35\textwidth-\offset,0-\offset) {(c)};
\end{tikzpicture}
\caption{Numerical solution of the wavenumber evolution problem~\eqref{eq:evolution} and \eqref{eq:constraint}, with stationary and uniform (in $\wno$-space) initial conditions.  (a) Mode amplitude $\coeff_{\wno}$ at time $\bt = 10^2$ (orange), $10^4$ (red), $10^6$ (purple), $10^8$ (blue) and $10^{10}$ (green) rescaled by the  maximum value of $\coeff_{\wno}$, ${\coeff_{\wno}}_\ast$, which is attained at $\wno=\wnostar$. (b) The hoop stress $\C = \stt{2}$ decays in time (blue points) but not quite in the way expected for a similarity solution, $\C\sim\bt^{-4/5}$ (black dashed line). (Note that the noise in the hoop stress prediction is an artefact of the numerics --- this results from changing the domain length to resolve the typical wavenumber at different times.) (c) The evolution of the dominant wavenumber $\wnostar$, defined as the value of $\wno$ at which the amplitude $\coeff_{\wno}$ is maximized (blue), together with a (black dashed) line showing the `expected' similarity behaviour $\wnostar\sim\bt^{-2/5}$. Inset: Plotting the rescaled dominant wavenumber, $\wnostar t^{2/5}$, on doubly logarithmic axes shows that the observed behaviour is \emph{not} self-similar. }
\label{fig:wavenumber}
\end{figure}
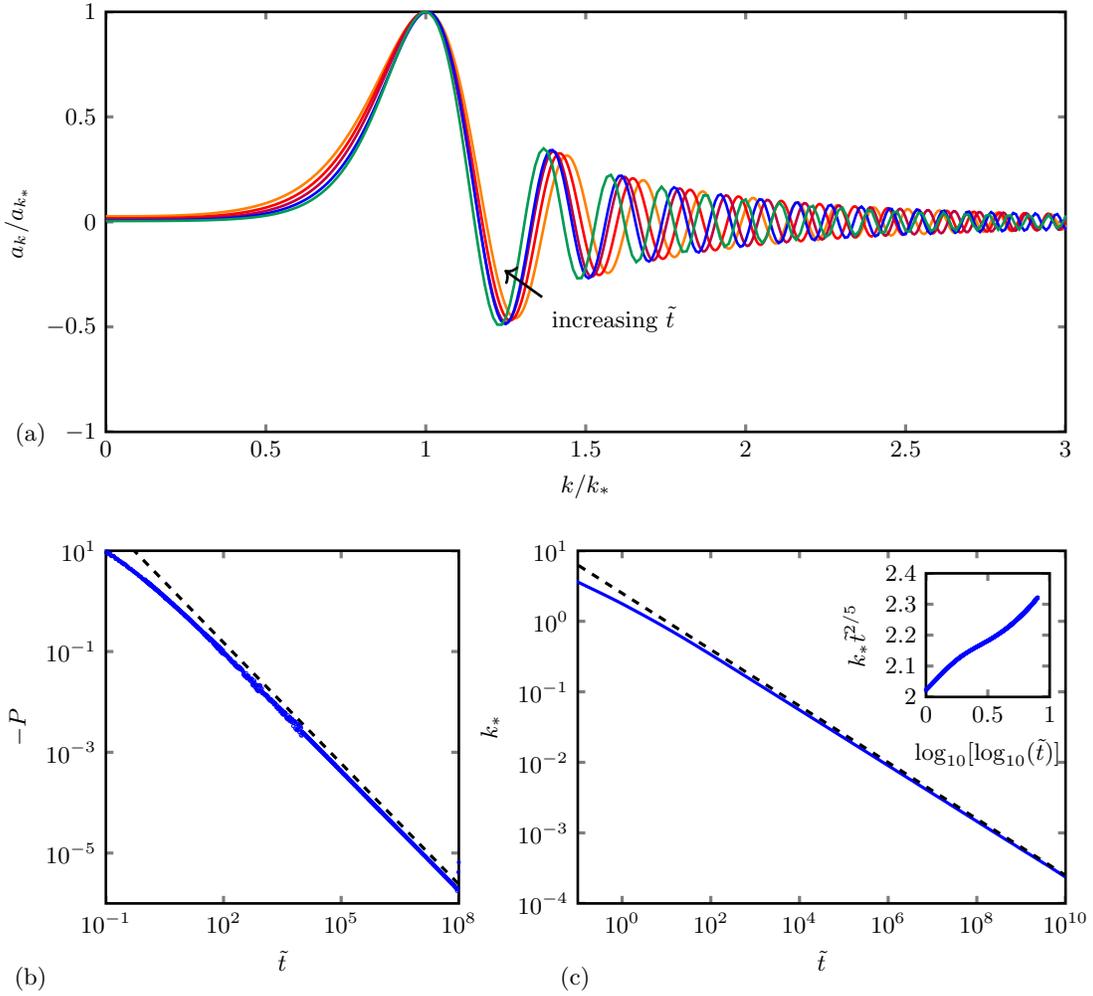

Snapshots of the amplitude $\coeff_{\wno}$ of the mode associated with the scaled wavenumber $\wno$ are shown in Figure~\ref{fig:wavenumber}(a): we observe that the amplitude varies smoothly with $\wno$,  is oscillatory, with a global maximum at some critical $\wno = \wnostar$, and decays as both $\wno \rightarrow 0$ (for infinitely long wavelengths) and $\wno \rightarrow \infty$ (for infinitesimally  short wavelengths); here we plot up to $\wno/\wnostar = 3$ only to clearly illustrate the behaviour for $\wno/\wnostar < 3$.  
 
As expected from the earlier discussion, our numerics show that the behaviour is \emph{not} self-similar: there is a consistent drift in the distribution $a_k(\bt)$ as $\bt$ evolves.  Furthermore, while the stress $\C = \stt{2} < 0$ is negative, and initially has a large magnitude that decays towards 0 as the sheet wrinkles (see Figure~\ref{fig:wavenumber}(b)), we note that the decay in $\C$ does not coincide with the $\bt^{-4/5}$ power law that would be expected of a similarity solution.

The dominant wavenumber decreases with time,  as shown in Figure~\ref{fig:wavenumber}(c); alternatively, the typical wavelength increases with time, in agreement with the experimentally-observed coarsening of wrinkles \cite{pnas}.  A dashed line corresponding to $\wnostar\propto \bt^{-2/5}$ (on logarithmic scales) is also shown in Figure~\ref{fig:wavenumber}(c) for comparison. This demonstrates that the  behaviour observed numerically is close to the power law $\bt^{-2/5}$ that would be expected from  the proposed similarity solution \eqref{eqn:SimSoln}.  However, given that the behaviour of the compressive stress $\C$ shows a systematic long-term drift from the expected power law,  we present in the inset of Figure~\ref{fig:wavenumber}(c) a ``compensated'' plot. This compensated plot shows a small but systematic drift from the scaling $\wnostar \sim \bt^{-2/5}$ over the time interval $\bt \in [1, 10^{10}]$.  In the related problem of viscosity-dominated wrinkle coarsening, an analogous deviation from power law behaviour is observed which is logarithmic and can be quantified~\cite{Kodio2017}.  The analysis carried out there cannot be adapted directly to this problem because of the second derivative in time arising from inertia in~\eqref{eq:evolution} as opposed to the first derivative that appears in the viscous case.  As a result, we are unable to go beyond numerical observations and speculation.  In that spirit we merely note that, for example the inset of Figure~\ref{fig:wavenumber}(c)  shows that $\wnostar \bt^{2/5}$ is approximately linear in $\log_{10}[\log_{10} \left( \bt \right)]$ at very late times.

\section{Comparison between static and dynamic wrinkling} \label{sec:comparison}
In this section we provide a brief discussion of the mechanisms underlying wrinkle coarsening, and compare the problem studied in this paper with static indentation. In static situations, the system selects the wavelength that minimizes the energy: the bending stiffness drives the system towards a large-amplitude (long wavelength) pattern, while substrate stiffness drives the system to adopt a small amplitude (short wavelength) pattern. The balance between these two effects yields the intermediate wavelength $\lambda \sim (B/K)^{1/4}$ given in \eqref{eqn:LambdaStatic}, which for quasi-statically indented floating sheets explains the wavelength observed in the flat portions of the sheet, far from the indenter \cite{Paulsen2016pnas}. In contrast, for the impact-induced wrinkling studied here, we have understood the evolving wrinkle formation not via energetic arguments but via pressure and force balances:
 the resistance to large-amplitude deformation comes 
at early times not from the hydrostatic pressure but rather the hydrodynamic pressure $p \sim \rho (\partial w/\partial t)^2$, and this restoring effect of fluid inertia can be reinterpreted as a dynamic substrate stiffness $K_{\mathrm{dyn}} \sim \rho \times \lambda/t^2$~\cite{pnas}.  Physically, the excess length imposed by radial displacement is absorbed by short-wavelength, small-amplitude wrinkles, since these minimize the amount of fluid that needs to be accelerated.  The wrinkles then coarsen as more fluid is accelerated.  

The energetic approach usually adopted in static problems and the force balance approach adopted in this dynamic situation can be reconciled using Hamilton's principle of stationary action \cite{Goldstein1980}. To confirm that the wrinkled solution considered in this paper is a `better' solution than the unwrinkled alternative, we consider the relevant energies in both the wrinkled and unwrinkled dynamic scenarios. 
{Considering first the action of the wrinkled response, the kinetic energy of the fluid dominates both the strain and bending energies provided the dimensionless time $t \gg t_{\text{edge}}$, with $t_{\mathrm{edge}}$ the time at which wrinkles reach the sheet edge  defined in~\eqref{eq:tedge}, and $t \gg \spar^5 (\rf/\lstar)^{5/2}$.  The second condition simply states that our model breaks down at sufficiently early times that the bending energy associated with the predicted wavelength becomes insurmountable.  (We also note that the hoop stress does not affect the elastic energy of the wrinkled response provided $\sigma_{\theta\theta} \ll \sigma_{rr}$, so this argument is not affected by a distinction between $\sigma_{\theta\theta} \sim \spar^2$ and $\sigma_{\theta\theta} \sim \spar^{3/2}$.) } {Comparing the magnitude of the action in the wrinkled case with that of the unwrinkled response (in which kinetic and stretching energies balance), we find that wrinkling is favourable provided that $t\gg t_{\mathrm{edge}}$, entirely consistent with our assumption of a fully wrinkled sheet made earlier.}

Another difference between static and dynamic wrinkle formation is the prediction that $\sigma_{\theta\theta} \sim \spar^{3/2}$ in some dynamic situations, as opposed to the $\sigma_{\theta\theta} \sim \spar^2$ scaling typical of fully-developed static wrinkles.  In particular, while we showed that the hoop stress in the wrinkled sheet is $O(\spar^2)$ everywhere on the $t = O(1)$ timescale over which the global deformation grows, and is $O(\spar^2)$ in the flat portion of the sheet even on the wrinkling timescale $t = O(\spar^{1/2})$, our model predicts that $\sigma_{\theta\theta} \sim \spar^{3/2}$ in the curved portion of the sheet on the wrinkling timescale.  In terms of the predictions made in this paper and their comparison with experiments, we have predicted (i) the propagation of the axisymmetric wave on the $t = O(1)$ timescale where $\sigma_{\theta\theta} \sim \delta^2$, and (ii) the evolution of wrinkles in the flat portion of the sheet.  The $t=O(1)$ global deflection, prediction (i), is not affected by the possibility that $\sigma_{\theta\theta} = O(\spar^{3/2})$, since our results rely only on $\sigma_{rr} \gg \sigma_{\theta\theta}$  
Our prediction for (ii), the wrinkles in the flat portion of the sheet, comes from a vertical force balance~\eqref{eq:evolution} and length constraint~\eqref{eq:constraint}.  We note that the excess length that drives wrinkling, see~\eqref{eq:constraint}, is generated by radial retraction of the sheet, which is driven by the vertical displacement of the inner portion of the sheet.  This could, in principle, by affected by changes in the vertical deflection of the sheet in $r = O(1)$ modifying the right-hand-side of~\eqref{eq:constraint}.

To check the accuracy of the excess length on the right-hand-side of~\eqref{eq:constraint}, we must {therefore} ask whether our prediction of the axisymmetric shape in the curved portion of the sheet, {on the $t = O(\spar^{1/2})$ wrinkling timescale}, is correct.  This axisymmetric behaviour is dependent only on the relaxation of the hoop stress, $\sigma_{\theta\theta}$ in comparison to the radial stress, $\sigma_{rr}$; the question we need to ask therefore is whether the hoop stress has collapsed to a sufficient extent to allow the radial stress $\sigma_{rr} = 1/r$. It is clear that a hoop stress $\sigma_{\theta\theta} \sim \spar^{3/2}$ is certainly small enough that the hoop stress may be considered to be relaxed in the way required for the emergence of the $\rm \sim t^{1/2}$ wave propagation observed experimentally~\cite{pnas}.  This adds further credence to our claim that the vertical displacement we predict for $r = O(1)$, $t = O(\delta^{1/2})$, and hence the right-hand-side of~\eqref{eq:constraint}, are correct.  Nonetheless, examining the curved portion of the sheet in the images in Figure~\ref{fig:snapshots} suggests that the wrinkle pattern in the curved portion of the sheet is different from that in the flat. While the behaviour of any wrinkles in this highly curved region is difficult to image experimentally, any differences in behaviour compared to that in the flat region might be explained by the emergence of different physical effects (see Appendix~\ref{app:wrinkles}) there.

\section{Conclusions}\label{sec:discussion}
In this paper, we have developed a mathematical model to describe impact onto a thin  elastic sheet floating on a liquid bath. In particular, we have focused on the early-time behaviour of the system, when the sheet is only moderately deformed and the impactor moves at constant speed.  We first addressed the simplest case, a point impactor, and then generalized our model to account for finite-radius spherical impactors.  Our analysis exploited the fact that the sheets of interest have very small bending stiffness, allowing us to focus on the ``high-bendability'' limit in which the resistance to bending is small compared to applied stresses (the limit $\spar\ll1$ with $\spar$ defined in equation \eqref{eqn:deltadefn}). Through an asymptotic analysis of the high-bendability limit, we were able to derive, from a single unified model, simplified models to describe (i) the transverse wave that propagates radially outward from the point of impact and (ii) the wrinkles that form as a result of the compressive azimuthal stress induced by impact.  In deriving this model, we found that these two phenomena occur on two different timescales: in the limit of high bendability, the wrinkles evolve on a timescale faster than the timescale associated with the transverse wave by a factor $\spar^{-1/2} \gg 1$.  This suggests that, in principle, the wrinkles should evolve quasi-statically as impact progresses.  However, in the experiments performed to date,  $\spar^{1/2}\approx0.3=O(1)$, so that, in reality, wrinkles  coarsen in tandem with the propagation of the transverse wave, as is observed experimentally \cite{pnas}.  

Our results explain some of the key differences between  impact on a floating elastic sheet and previous results for either impact on water or (quasi-static) indentation of floating sheets. For the propagating transverse wave, we found a similarity solution in which  the distance of propagation scales with $t^{1/2}$, rather than the $t^{2/3}$ behaviour typical of capillary-driven flows \cite{KellerMiksis1983,VellaMetcalfe2007}. This difference arises from the spatially varying  tension within the sheet, which is, in turn, a consequence of wrinkling \cite{Davidovitch2011,Vella2015PRL}.  Moreover, this $t^{1/2}$ behaviour is also different from the results of classic Wagner impact theory~\cite{Philippi2016,Wagner1932,Korobkin1988} since (i) the tension of the interface is accounted for and (ii) the radius of the impactor cannot be scaled out of the problem, but rather remains as a single dimensionless parameter that affects the pre-factor in the $t^{1/2}$ scaling. We also saw that a higher-order correction to our early-time similarity solution suggests that the wave spreads out more rapidly at later times as nonlinear effects come into play; this is in qualitative agreement with experimental observations \cite{pnas}.  

The second focus of this paper has been on the early-time evolution of the wrinkle wavelength during impact: while quasi-static indentation experiments have shown that the number of wrinkles increases (wavelength decreases) with increasing indentation, here 
fluid inertia opposes the formation of large wavelength wrinkles, and so the sheet at first forms many fine wrinkles that subsequently coarsen.  The observed wrinkle wavelength is therefore truly dynamic, and not simply quasi-statically evolving with impact depth. A simple scaling analysis suggests that the typical wavenumber of the wrinkle pattern $\wnostar\sim t^{-2/5}$. However, numerical solution of the governing equations suggests that this coarsening is slowed slightly by the geometric constraint that the compression arising from the radial displacement of the sheet must be exactly accommodated by the arc-length along wrinkles.  We are not able to explain this deviation from power law behaviour quantitatively, but note that this problem offers an inertial analog of the constraint-limited coarsening of viscously-damped wrinkles \cite{Kodio2017}, for which a logarithmic correction to power-law coarsening was derived analytically.

There are some limitations on the validity of our modelling approach at early times.  Firstly, the model presented in this paper breaks down at sufficiently early times ($<0.1$ms, see the first panel of Figure~\ref{fig:snapshots}) that the wrinkles do not yet cover the entire sheet, and the wrinkle length is still evolving.  As a result, it is not appropriate to compare our work to the observations of~\cite{BugraToga2013} on the concurrent evolution of wrinkle length and number in geometries where wrinkles form in an annulus.  An second limitation is that the hoop stress, which diverges $\sim \delta^2 \bt^{-4/5}$ at early times, must have undergone significant collapse before our model is valid.    Finally, we note that mathematical modelling of the very early-time dynamics of the wrinkling sheet considered in this study are complicated by the inertia of the underlying liquid, and the associated separation of timescales.  An alternative study of a dynamic wrinkling configuration without an underlying fluid may allow for further theoretical progress.

We did not address some of the features of this system that become apparent at later times. For example, further investigation is required to determine how the system transitions from the balance between bending stiffness and hydrodynamics that is important early on to the static balance between bending, surface tension and hydrostatic pressure within the liquid that governs the later behaviour \cite{Paulsen2016pnas} if the sphere is ultimately halted by the sheet. Indeed, the transition between the trapping of the sphere at the interface and impact causing the sheet to be dragged under water has not been explored, since we have  assumed throughout that the sphere moves at a constant downward velocity.  This floating/sinking transition may be important not only in the suppression of splashing \cite{poopsplash}, but also as an alternative route for the controlled encapsulation of droplets within an elastic sheet \cite{Kumar2018}.

\acknowledgments

This research was supported by the European Research Council under the European Horizon 2020 Programme, ERC Grant Agreement no.~637334 (DV).  We are grateful to Alfonso Castrej\'{o}n-Pita for the loan of equipment used to make Figure 1.  We also wish to thank an anonymous referee for detailed and though-provoking comments.

\appendix
\section{Correction to axisymmetric similarity solution}\label{sec:simsolcorr}
In this Appendix we give further information on the leading-order axisymmetric surface profile calculated in \S\ref{sec:elastocapillarywave}.  There, we considered an early-time limit, neglecting non-linear terms in the boundary conditions~\eqref{eq:kinematic-lo-a} and~\eqref{eq:dynlo-a}, and thus calculated a self-similar interface shape of the corresponding wave that propagates radially outward.    
This leading-order behaviour gives a good account of experimental observations, but slightly under-predicts the speed of the propagating wave.  We now assess the increasing importance of the nonlinear terms in~\eqref{eq:kinematic-lo-a} and \eqref{eq:dynlo-a} as time increases.  In \S\ref{sec:simsol} we wrote an early-time self-similar ansatz~\eqref{eq:selfsim}.  Strictly speaking we should write an asymptotic expansion of the form
\begin{subequations}
\begin{equation}
\w{0}(r,t) = t \Hsim(\xi) + t^{3/2} H_1(\xi) + \cdots,
\end{equation}
\begin{equation}
\p{0}(r,z,t) = t^{1/2} \f(\xi,\zeta) + t f_1(\xi,\zeta) + \cdots,
\end{equation}
\begin{equation}
\rca = t^{1/2} \con_0 + t \con_1 + \cdots
\end{equation}
\end{subequations}
in powers of the small parameter $t \ll 1$.  The leading-order variables $\f$, $\Hsim$ are then calculated in the main text, and we now determine the terms $H_1(\xi)$ etc., and investigate whether they might account for the observed discrepancies between our predictions and experiments.  

The correction to the velocity potential satisfies
\begin{equation}
\frac{1}{\xi} \frac{\partial}{\partial \xi} \left( \xi \frac{\partial \fcor}{\partial \xi} \right) + \frac{\partial^2 \fcor}{\partial \zeta^2} = 0,
\end{equation}
with kinematic boundary condition
\begin{equation}
\frac{\partial \fcor}{\partial \zeta} = \left\{ \begin{array}{cc}
- \Hsim \tfrac{\partial^2 \f}{\partial \zeta^2} + \frac{\xi}{\simpar} \frac{\partial \f}{\partial \xi}, & \xi \leq \con_0
\\
\left(\tfrac{3}{2} \Hcor - \tfrac{1}{2} \xi \tfrac{\mathrm{d} \Hcor}{\mathrm{d} \xi}\right) - \Hsim \frac{\partial^2 \f}{\partial \zeta^2} + \tfrac{\mathrm{d} \Hsim}{\mathrm{d} \xi} \tfrac{\partial \f}{\partial \xi}, & \xi > \con_0
\end{array} \right. ,
\end{equation}
and  dynamic boundary condition
\begin{equation} \label{eq:simdyn-cor}
\left( \fcor - \frac{1}{2}\xi \frac{\partial \fcor}{\partial \xi} \right) - \frac{1}{\xi} \frac{\mathrm{d}^2 \Hcor}{\mathrm{d} \xi^2} = \frac{1}{2} \xi \Hsim \frac{\partial^2 \f}{\partial \zeta \partial \xi} - \frac{1}{2} \left[ \left( \frac{\partial \f}{\partial \xi}\right)^2 + \left( \frac{\partial \f}{\partial \zeta} \right)^2 \right], \quad \xi > \con_0
\end{equation}
on $\zeta = 0$.  There is no correction to the surface height inside the contact region, so $\Hcor = 0$ for $\xi \leq \con_0$.  The conditions at $\xi = 0$ and in the far field are simply
\begin{equation}
\frac{\partial \fcor}{\partial \xi} = 0, \quad \xi = 0,
\end{equation}
and
\begin{subequations} \label{eq:farfield1}
\begin{equation}
\Hcor \rightarrow 0 \quad \text{as} \quad \xi \rightarrow \infty,
\end{equation}
\begin{equation}
\fcor \rightarrow 0 \quad \text{as} \quad \zeta^2 + \xi^2 \rightarrow \infty.
\end{equation}
\end{subequations}
These equations may be solved numerically using second-order finite differences, as described in \S\ref{sec:simsol} for the leading-order problem.  The correction to the location of the contact line may then be inferred from the condition of smoothness of the interface at the contact point, which is transformed to
\begin{equation}\label{eq:contactcorrection}
\con_1 = \frac{ \tfrac{\mathrm{d} \Hcor}{\mathrm{d} \xi} \left( \con_0 \right)}{\tfrac{1}{\simpar} - \tfrac{\mathrm{d}^2 \Hsim}{\mathrm{d} \xi^2} \left( \con_0 \right)}.
\end{equation}
Similarly, a correction to the location of the minimum may be determined via Taylor expansion, to give an expression equivalent to~\eqref{eq:contactcorrection}, without the $1/\simpar$ since the slope is zero at the minimum.  A two-term prediction for the location of the first minimum is shown in Figure~\ref{fig:similarity+expts}(a).  
This correction to the leading-order similarity solution suggests that the nonlinear terms cause the wave to spread out slightly more rapidly than the leading-order linear system suggests, in qualitative agreement with observations.  

\section{Impactor deceleration} \label{sec:deceleration}
Throughout this paper we have assumed that the speed of the impactor is constant, i.e.\ that impact does not slow the impactor down significantly during the time interval of interest.   To assess the validity of this assumption, and hence of our early-time similarity solution, we now estimate the deceleration of the sphere during the early stages of impact.  

\begin{figure}[ht]
\begin{tikzpicture}

\draw[circle,thick,ForestGreen] (0,-0.63) circle (0.9cm);

\node[blue] at (-4.0,-1.0) {$\text{hydrostatic pressure} \sim - \rho \frac{\partial \varphi}{\partial t}$};
\node[red] at (3.3,-1.5) {$\text{interfacial pressure jump} \sim \sigma_{rr} \frac{\partial^2 w}{\partial r^2}$};
\draw[->,ForestGreen] (0.0,-1.0) to (0.0,-2.0) node[below,ForestGreen] {gravity};

\begin{axis}[
    at = {(0,0)}, anchor = {origin},
    width = 0.6\textwidth,
    height = 0.2\textwidth,
    xmin = -3, xmax = 3,
    ymin = -1.01, ymax = 0.3,
    every axis plot/.append style={thick},
    axis line style={draw=none},
    ticks=none,
    ]
    \addplot[black] table {similarity-solution/intro/profile_epsilon2.txt};
    \addplot[red,very thick] table {similarity-solution/intro/contact_epsilon2.txt};
\end{axis}

\end{tikzpicture}
{\caption{The net force acting on the impacting sphere arises from gravity, tension in the deformed elastic sheet, and hydrodynamic pressure in the underlying liquid.
}
\label{fig:deceleration}}
\end{figure}
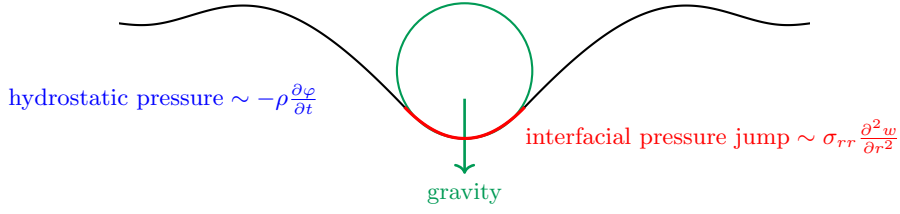

During impact, the impactor experiences an upward force as a result of the tension in the stretched sheet and hydrodynamic pressure in the underlying fluid, as illustrated in Figure~\ref{fig:deceleration}, as well as the downward pull of gravity.  
In dimensionless variables, Newton's second law applied to the sphere gives in dimensionless variables
\begin{equation}\label{eq:dimforce}
\left( \frac{4 \pi}{3} \rs^3 \rhosphere \right) \frac{V^2}{\lstar} \frac{\mathrm{d}^2 z}{\mathrm{d} t^2} = \int_0^{2\pi} \int_0^{r_c(t)} \left( \frac{\gamma \rf}{\lstar^2} \frac{1}{r} \frac{\partial^2 w}{\partial r^2} - \rho \V^2 \frac{\partial \varphi}{\partial t} \right)\ \lstar^2\ r\ \mathrm{d} r\ \mathrm{d} \theta - \frac{4 \pi}{3} \rs^3 \rhosphere g, 
\end{equation} 
where $\rhosphere$ is the density of the spherical impactor, $z$ is the (dimensionless) height of its centre of mass and the right-hand-side consists of the vertical force from the tension within the sheet and hydrodynamic pressure (both of which only act on the sphere within the contact region), together with the impactor's weight ($g$ being the acceleration due to gravity).  Applying the early-time similarity solution described in \S\ref{sec:simsol}, this can be rewritten as
\begin{equation}
\frac{\mathrm{d}^2 z}{\mathrm{d} t^2} = \frac{1}{\simpar{^3}} \frac{\rho}{\rhosphere} \left[ \frac{\con_0}{\simpar}  - \frac{1}{2} \int_0^{\con_0} \left(\f - \xi \frac{\partial f_0}{\partial \xi} \right) \xi\ \mathrm{d} \xi \right] \frac{3}{2} {t}^{1/2} - \frac{g\lstar}{\V^2}.
\end{equation} 
Evaluating the similarity solution, we find that the integral quantifying hydrodynamic pressure is negative, so the term in square brackets is $O(1)$ and positive.  Integrating, we find
\begin{equation}
z = \alpha - t + \frac{2}{5}  \frac{1}{\simpar{^3}} \frac{\rho}{\rhosphere} \left[ \frac{\con_0}{\simpar}  - \frac{1}{2} \int_0^{\con_0} \left(\f - \xi \frac{\partial f_0}{\partial \xi} \right) \xi\ \mathrm{d} \xi \right] {t}^{5/2} - \frac{g\lstar}{2 \V^2} t^2.
\end{equation} 
For the experiments reported in~\cite{pnas}, the prefactor of  the $-t^{5/2}$ (impact-induced deceleration) term is $O(10^{-1})$ and the prefactor of the $t^2$ (gravitational acceleration) term is $O(10^{-3})$--$O(10^{-2})$.  As a result, the impact speed is indeed approximately constant for $t < 1$, This suggests that the most important restriction on the validity of the model presented in this paper is the requirement that the contact region be small in comparison to the sphere's radius (allowing us to approximate the contacting portion of the sheet by a parabloid, as discussed in \S\ref{sec:simsol}).

We note that for smaller values of $\alpha$ (i.e.~smaller or slower spheres, or larger sheet radii), the prefactor of $t^{5/2}$ would be larger in magnitude, and deceleration would be significant.  In this scenario, the similarity solution~\eqref{eq:laplace4}--\eqref{eq:farfield} will break down once
\begin{equation} \label{eq:app:decelerationtime}
t \sim \simpar^2 \left\{ \frac{2}{5}  \frac{\rho}{\rhosphere} \left[ \frac{\con_0}{\simpar}  - \frac{1}{2} \int_0^{\con_0} \left(\f - \xi \frac{\partial f_0}{\partial \xi} \right) \xi\ \mathrm{d} \xi \right] \right\}^{-2/3},
\end{equation} 
beyond which the model must be modified to account for the fact that the impact depth does not increase linearly with time, but rather is determined by a coupled force balance of the form~\eqref{eq:dimforce}.  Furthermore, in such cases, the upward force on the sphere may be sufficient for it (and the sheet) to rebound, or the sheet maybe become submerged below the interface.

\section{Wrinkles}\label{app:wrinkles}
\subsection{Rescalings}
The dominant balance argument presented in \S\ref{sec:nondim-action} suggests that the natural timescale on which to study wrinkle formation is $\bt = \spar^{-1/2} t = O(1)$.  In this Appendix, we study the governing equations~\eqref{eq:laplace2}--\eqref{eq:farfieldall} on this shorter timescale, and derive a simplified model for the inertia-modulated evolution of wrinkles in the floating elastic sheet.   Our analysis on the $t = O(1)$ timescale suggests that the disturbance induced by the impact will grow in time, and so is small when $t = \spar^{1/2} \bt \ll 1$.  Motivated by this observation, we rescale the axisymmetric variables according to
\begin{subequations} \label{eq:shortscalings}
\begin{equation}
\bar{w}(r,t) = \spar^{1/2} \bt H(\xi,t), \quad \bar{\varphi}(r,z,t) = \spar^{1/4} \bt^{1/2} f(\xi,\zeta,\bt), \quad \text{where } \xi = \frac{r}{\spar^{1/4} \bt^{1/2}}, \quad \zeta = \frac{z}{\spar^{1/4} \bt^{1/2}}.
\end{equation}

Similarly, the balance of terms in the length constraint~\eqref{eq:lengthquasi} and boundary conditions~\eqref{eq:kinematic} and \eqref{eq:vertforce} motivates rescaling the oscillatory components according to
\begin{equation}
\R{w}(r,t) = \spar^{11/8} W(r,\btheta,\bt), \quad \R{\varphi} = \spar^{15/8} \Phi(r,\btheta,z,\bz,\bt).
\end{equation}
\end{subequations}

On this short timescale, where the axisymmetric and oscillatory components of variables scale differently, we shall see that it is easiest to consider the axisymmetric and oscillatory components of each equation separately from the outset.  When studying the axisymmetric components, anticipating self-similar behaviour, we write
\begin{equation}
\bar{\sigma}_{rr} = \spar^{-1/4} \bt^{-1/2} S_{rr}.
\end{equation}
For the oscillatory components of the stress tensor, examining the horizontal force balance~\eqref{eq:horzforce} and stress--strain relationships~\eqref{eq:stressstrain} suggests that 
\begin{equation}
\R{\sigma}_{rr} \sim \spar^{13/8}, \quad \R{\sigma}_{r\theta} \sim \spar^{21/8} , \quad \R{\sigma}_{\theta\theta} \sim
\spar^{29/8},
\end{equation}
while we anticipate that the in-plane displacement components
\begin{equation}
\bar{u}_r \sim \spar^{3/4}, \quad \R{u}_r \sim \spar^{13/8}, \quad \bar{u}_{\theta} \equiv 0, \quad \hat{u}_{\theta}\sim \spar^{14/8}.
\end{equation}

We now examine the vertical force balance~\eqref{eq:vertforce} in detail.  Starting with the non-axisymmetric component, we observe that the term associated with hoop stress appears first, and so the hoop stress is, at most, $\sigma_{\theta\theta} \sim \spar^{3/2}$.  We can then solve the axisymmetric horizontal force balance to find
\begin{equation}
\bar{\sigma}_{rr} = \frac{1}{r} + O(\spar^{3/2}), \quad S_{rr} = \frac{1}{\eta} + \mathrm{h. o. t.}, \quad \bar{\sigma}_{r\theta} \equiv 0.
\end{equation}

\subsection{Axisymmetric behaviour}
Rescaling the axisymmetric components of the governing equations~\eqref{eq:laplace2}--\eqref{eq:farfieldall} for $t = \spar^{1/2} \bt$ and evaluating at leading order, we find that
\begin{equation}
\frac{1}{\xi} \frac{\partial}{\partial \xi} \left( \xi \frac{\partial \f}{\partial \xi} \right) + \frac{\partial^2 \f}{\partial \xi} = 0,
\end{equation}
with boundary conditions
\begin{subequations}
\begin{equation}
\frac{1}{2} \left[ t f_{0,t} + \f - \xi f_{0{,}\xi} \right] = \frac{1}{\xi} \Hsimf_{0,\xi \xi},
\end{equation}
\begin{equation}
\ff_{0{,}\zeta} = t \Hsimf_{0{,}t} + \Hsim - \frac{1}{2} \xi \Hsimf_{0{,}\xi},
\end{equation}
on $\zeta = 0$, and
\begin{equation}
\Hsim = -1, \quad \ff_{0{,}\xi} = 0 \quad \text{at} \quad \xi = 0,
\end{equation}
\begin{equation}
\f \rightarrow 0 \quad \text{as} \quad \xi^2 + \zeta^2 \rightarrow \infty,
\end{equation}
\begin{equation}
\Hsim \rightarrow 0 \quad \text{as} \quad \xi \rightarrow \infty.
\end{equation}
\end{subequations}
This is satisfied by the same solution $\Hsim(\xi)$, $\f(\xi,\zeta)$ as the linearized problem outlined in~\S\ref{sec:elastocapillarywave}, as expected.  The corresponding axisymmetric in-plane displacements are
\begin{equation} \label{eq:app:retraction}
\bar{u}_r \approx - \frac{1}{2} \spar^{3/4} \bt^{3/2} \int_{0}^{\infty} \left(\frac{\mathrm{d} H_0}{\mathrm{d} \xi} \right)^2 \mathrm{d} \xi, \quad \bar{u}_{\theta} \equiv 0.
\end{equation}

\subsection{Wrinkling behaviour}
We now turn our attention back to the oscillatory component of the vertical force balance~\eqref{eq:vertforce}.  Rescaling as outlined in~\eqref{eq:shortscalings} and evaluating at leading order, we find
\begin{equation}
\frac{\partial \f}{\partial \zeta} \frac{\partial \Phi_0}{\partial \bz} = \bar{\sigma}_{\theta\theta}^{(3/2)} \frac{1}{r^2} \frac{\partial^2 W_0}{\partial \btheta^2}.
\end{equation}
This suggests that if the leading-order axisymmetric and oscillatory velocity potential components are both non-zero, then there is a non-zero hoop stress at $O(\spar^{3/2})$, which is \emph{larger} than the $O(\spar^2)$ residual hoop stress normally observed in static, high-bendability wrinkling problems~\cite{Davidovitch2011}.   We focus our attention on the region $r \gg 1$ where the sheet is flat, so that $\partial \f/\partial \zeta = 0$, and hence $\stt{3/2} = 0$.  We can then write down a complete problem for the leading-order oscillatory velocity potential and displacement, namely the velocity potential is governed by a Laplace equation~\eqref{eq:laplace2}
\begin{equation} \label{eq:finalw}
\frac{1}{r^2} \frac{\partial^2 \Phi_0}{\partial \btheta^2} + \frac{\partial^2 \Phi_0}{\partial \bz^2} = 0,
\end{equation}
The boundary conditions at the free surface~\eqref{eq:kinematic}, \eqref{eq:vertforce} are transformed to
\begin{equation} \label{eq:finalvert}
\frac{\partial \Phi_0}{\partial \bz} = \frac{\partial W_0}{\partial \bt}, \quad \frac{\partial \Phi_0}{\partial {\bt}} + \frac{1}{r^4} \frac{\partial^4 W_0}{\partial \btheta^4} = \sigma_{\theta\theta}^{(2)} \frac{1}{r^2} \frac{\partial W_0}{\partial \btheta^2}
\end{equation}
on $\bz = 0$, where the terms associated with tension- and curvature-induced stiffness seen in the quasistatic problem~\eqref{eq:wquasi} are neglected in the limit $r \gg 1$.  The velocity potential $\Phi_0$ must decay in the far field $\bz \rightarrow - \infty$, and the entire system must be periodic in $\btheta$.  Finally, the unknown hoop stress $\bar{\sigma}_{\theta\theta}^{(2)}$ is determined by the length constraint
\begin{equation} \label{eq:finalconstraint}
\overline{ \left[ \left( \frac{1}{r} \frac{\partial W_0}{\partial \btheta} \right)^2 \right] } = \frac{1}{r} \bt^{3/2} \int_0^{\infty} \left( \frac{\mathrm{d} H_0}{\mathrm{d} \xi} \right)^2\ \mathrm{d} \xi
\end{equation}
for a point indenter, or the finite-indenter equivalent~\eqref{eq:retraction2}.
The forces governing the dynamic wrinkling of the sheet are therefore the compressive azimuthal force [the term on the right-hand side of~\eqref{eq:finalvert}] and two different terms that resist this compression (from left to right on the left-hand side: the inertia of the underlying fluid and the bending stiffness of the elastic sheet). 

\subsection{Fourier transform}
The reduced system of equations~\eqref{eq:finalw}--\eqref{eq:finalconstraint} governing the leading-order oscillatory behaviour are satisfied by solutions of the form
\begin{subequations} \label{eq:fourier-app}
\begin{equation}
\Phi_0(r\btheta,\bz,\bt) = \int_{-\infty}^{\infty} \frac{1}{|\wno|} \frac{\mathrm{d} \coeff_{\wno}}{\mathrm{d} \bt} \mathrm{e}^{i \wno \btheta r} \mathrm{e}^{|\wno| \bz}\ \mathrm{d} \wno,
\end{equation} 
\begin{equation}
W_0(r\btheta,\bt) = \int_{-\infty}^{\infty} \coeff_{\wno} \mathrm{e}^{i \wno \btheta r}\ \mathrm{d} \wno.
\end{equation}
\end{subequations}
with $\Phi_0 \rightarrow 0$ as $\bz \rightarrow -\infty$ as required, and the coefficients $a_{\wno}(r,\bt)$ to be determined.   The vertical force balance \eqref{eq:finalvert} is transformed to
\begin{equation} \label{eq:evolution-app}
\frac{\mathrm{d}^2 \coeff_{\wno}}{\mathrm{d} \bt^2} = - |\wno| \left[ \wno^4 - \stt{2}(t) \wno^2 \right] \coeff_{\wno},
\end{equation}
The length constraint~\eqref{eq:finalconstraint} must also be rewritten: we note that $a_{\wno}$ can be scaled to absorb any prefactor in the edge retraction, so that the dominant wavenumber $\wno_{\ast}(t)$ at which $a_{\wno}$ is maximized is only affected by the form of the power law.  It is therefore sufficient to solve~\eqref{eq:evolution-app} subject to the constraint 
\begin{equation} \label{eq:constraint-app}
\int_{-\infty}^{\infty} \wno^2 |\coeff_{\wno}|^2\ \mathrm{d} \wno = \bt^{\hspace{0.5mm} 3/2},
\end{equation}
and appropriate initial conditions, for example supplying the amplitude of $\coeff_{\wno}$ and $\partial \coeff_{\wno}/\partial \bt$ at $\bt = 0$.  This is the approach adopted for comparison with experimental measurements in~\S\ref{sec:wrinklexpts}.  However, in~\S\ref{sec:wrinklenumber} of this paper our priority is to determine whether an evolving wrinkle profile should be self-similar, and how it might deviate from power-law behaviour, so we focus on the simpler case of constant radial retraction and set the right-hand side of~\eqref{eq:constraint-app} to 1.

\bibliography{refs}        

\end{document}